\documentclass[11pt,preprint,superscriptaddress,aps,showkeys,nofootinbib]{revtex4}
\usepackage{amssymb,amsmath,amsfonts}
\usepackage{graphicx}
\usepackage{graphics}
\usepackage{eepic,epsfig}
\usepackage{indentfirst}
\usepackage[utf8]{inputenc}
\usepackage[T1]{fontenc}

1.60cm \makeatletter \@addtoreset{equation}{section}

\makeatother

\begin{document}
\title{Thermal corrections to the Casimir energy in a Lorentz-breaking scalar field theory}
\author{M. B. Cruz}
\affiliation{Departamento de F\'{\i}sica, Universidade Federal da Para\'{\i}ba\\
 Caixa Postal 5008, 58051-970, Jo\~ao Pessoa, Para\'{\i}ba, Brazil}
\email{messiasdebritocruz@gmail.com, emello, petrov@fisica.ufpb.br}
\author{E. R. Bezerra de Mello}
\affiliation{Departamento de F\'{\i}sica, Universidade Federal da Para\'{\i}ba\\
 Caixa Postal 5008, 58051-970, Jo\~ao Pessoa, Para\'{\i}ba, Brazil}
\email{messiasdebritocruz@gmail.com,  emello, petrov@fisica.ufpb.br}
\author{A. Yu. Petrov}
\affiliation{Departamento de F\'{\i}sica, Universidade Federal da Para\'{\i}ba\\
 Caixa Postal 5008, 58051-970, Jo\~ao Pessoa, Para\'{\i}ba, Brazil}
\email{messiasdebritocruz@gmail.com,  emello, petrov@fisica.ufpb.br}


\begin{abstract}

In this paper, we investigate the thermal effect on the Casimir energy associated with a massive scalar quantum field confined between two  large parallel plates in a $CPT$-even, aether-like Lorentz-breaking  scalar field theory. In order to do that we consider a nonzero chemical potential for the scalar field  assumed to be in thermal equilibrium at  some finite temperature. The calculations of the energies are developed by using the Abel-Plana summation formula, and the corresponding results are analyzed in several asymptotic regimes of the parameters of the system, like mass, separations between the plates and temperature.

\end{abstract}

\maketitle

\newpage
\section{Introduction} \label{introduction}

The Casimir effect is a macroscopic manifestation of the quantum vacuum. It was proposed in $1948$ by H. B. Casimir in 1948 \cite{Casimir:1948dh}, and experimentally confirmed ten years later  by M. J. Sparnaay \cite{Sparnaay:1958wg}. Thus it has being recognized as one of the most interesting phenomena related to quantum vacuum fluctuations.

The more common case for the  study of Casimir effect is the scenario of interaction between two parallel plates in a quantum vacuum. In fact, as  a consequence of the quantization of the wavelength in the direction perpendicular to the plates, an interaction between two uncharged parallel conductors plates placed in the  vacuum is generated. Thus, the Casimir effect is a phenomenon of purely quantum nature. In a general way, we can define the Casimir effect as a pressure (force per unit area) when boundary conditions are imposed on the quantum fields on the plates. In fact, the imposition of boundary condition on the quantum fields alters the zero-point fluctuation spectrum and result in additional shift in the vacuum expectation value of the energy density operator.

Specifically  Casimir investigated the quantization of the electromagnetic field confined between two uncharged parallel conductors. In general, the electromagnetic field  includes all possible wavelengths, and after placing the two plates, due to boundary  conditions, only some wavelengths can persist between them. When the quantum vacuum energy is  computed, an infinite amount is found, so one faces the problem of the ultraviolet divergence. To solve this problem we use  the mechanism of renormalization, that is, we  subtract the energy of free quantum vacuum from the energy between plates, arriving at a finite energy.

In recent years, the violation of Lorentz symmetry in quantum field theories (QFTs), has been questioned both in the theoretical and experimental context. In \cite{Kostelecky:1988zi}, V. A. Kostelecky and S. Samuel described a mechanism in string theory that allows the violation of Lorentz symmetry at the Planck energy scale. In this publication, the Lorentz symmetry is spontaneously broken due the emergence of preferential direction in the space-time induced by a non-vanishing vacuum expectation value of some component of vector and tensor fields. Consequently, if there is a violation of the Lorentz symmetry at the Planck energy scale in a more fundamental theory, the effects of this breakdown must manifest itself in other energy scales in different QFT models. In this way in \cite{ColKost}, it was  investigated the consequence of a general Lorentz violation symmetry in the context of Standard Model, including CPT-even and CPT-odd terms. Other mechanisms of violation of Lorentz 
symmetry are possible, such as space-time non-commutativity \cite{Carroll:2001ws, Anisimov:2001zc, Carlson:2001sw, Hewett:2000zp, Bertolami:2003nm}, variation of coupling constants \cite{Kostelecky:2002ca, Anchordoqui:2003ij, Bertolami:1997iy}  and modifications  of quantum gravity \cite{Alfaro:1999wd,Alfaro:2001rb}.  

 The anisotropy of the spacetime in a model of QFT resulting from a  Lorentz symmetry breaking, modifies the spectrum of the Hamiltonian operator and consequently the dispersion relation. At the same time, it should be noted that one of the best studied quantum effects, considered both theoretically and experimentally, is the Casimir effect.
Therefore it can be naturally treated as an excellent laboratory for the study of Lorentz symmetry violation. 

The first studies  of the Casimir effect in the context of Lorentz symmetry breaking were carried out in quantum electrodynamics (QED) 
\cite{Frank:2006ww,Kharlanov:2009pv,Silva:2016laz}. This naturally calls interest to these studies within other field theory models.

In the zero temperature case, the Casimir effect produced by  a massless scalar quantum field in theory where the Lorentz symmetry is broken in a strong,  Horava-Lifshitz-like manner, was analyzed in \cite{Ferrari:2010dj} and \cite{Ulion:2015kjx}. Moreover, for the massive scalar quantum  field, in a Lorentz-breaking scenario introduced by direct coupling between a constant vector and the derivative of the field, 
the Casimir effect was considered in \cite{Cruz:2017kfo};. There the main objective  was to investigate how the violation of the Lorentz symmetry, codified by a parameter $\lambda$, modifies the Casimir energy. Another point that deserves to be analyzed is influence of a nonzero temperature on the scalar Casimir effect.

The thermal corrections for Casimir effect (considering the case of the electromagnetic field) were first calculated by  Lifschitz \cite{Lifshitz:1956zz}, Fierz \cite{Fierz:1960zq} and Sauer \cite{Sauer:1962}. Posteriorly, Mehra 
\cite{Mehra:1967wf}, Boyer \cite{Boyer:1969xy}, Brown \cite{Brown:1969na} and Schwinger \cite{schwinger1975casimir} calculated  the thermal corrections using different techniques. Later Ambjorn \cite{Ambjorn:1981xw} and Konolish 
\cite{konoplich1989casimir} found the thermal corrections for the Casimir effect generated by a massless scalar field, in  a space-time with $D$ dimensions, in the  low temperature limit.

In this article, we extend our previous analysis \cite{Cruz:2017kfo}, considering a massive scalar field with a nonzero chemical potential, $\mu$, in thermodynamic equilibrium with a thermal bath possessing  a finite temperature, $T=\beta^{-1}$. The Casimir energy associated with this configuration is calculated in several asymptotic regimes of the physical parameters of the system. 

The structure of the paper is as  follows. In the section 2, we present basic definitions necessary for our studies. In  the section 3, we calculate the thermal modifications of the Casimir effect. Our results are summarized in the section 4. The  sections 5 and 6 are Appendices involving intermediate steps of our calculations. Here, we will use the natural units $c=\hbar=k_{B}=1$ and signature $-2$ for the Minkowski metric tensor.

\section{Basic definitions} \label{basic_definitions}

Here, in this section, we introduce the theoretical model that we want to investigate. This model is described by a massive real  scalar quantum field, $\phi(x)$, whose dynamics is governed by a $CPT-$even Lorentz-breaking extension of the  scalar field Lagrangian below\footnote{Originally, it was introduced as an 
ingredient of the Lorentz-violating extension of the standard model \cite{ColKost}.}:
\begin{eqnarray}
\label{lagrangian_density}
{\cal{L}}=\frac{1}{2}\left[\partial_\mu\phi\partial^\mu\phi+{\lambda} (u\cdot\partial\phi)^2+m^2\phi^2\right] \ .
\end{eqnarray} 
The Lorentz symmetry violating term is introduced through the direct coupling between the derivative of the scalar field with a fixed dimensionless constant vector $u^{\mu}$ \footnote{At the quantum level, this model for the scalar field was considered in Ref. \cite{Gomes:2009ch}.},  with $u^{\mu}u_{\mu}$ is equal either to $\pm 1$ or to 0.  Here the dimensionless parameter $\lambda$ is supposed to be much smaller that one.  It codifies the scale of the Lorentz symmetry violation.

Varying the Lagrangian density above with respect to the scalar field, we get
\begin{eqnarray}
\label{klein_gordon_equation}
\left[\Box+\lambda(u\cdot\partial)^2+m^2\right]\phi(x)=0 \ . 
\end{eqnarray}
Moreover, the energy-momentum tensor, as usual, is defined as:
\begin{eqnarray}
T^{\mu\nu}&=&\frac{\partial{\cal{L}}}{\partial(\partial_\mu\phi)}(\partial^\nu\phi) -\eta^{\mu\nu}{\cal{L}} \ ,
\end{eqnarray}
and consequently we find,
\begin{eqnarray}
\label{energy-tensor}
T^{\mu\nu}&=&(\partial^\mu\phi)(\partial^\nu\phi)+\lambda u^\mu(\partial^\nu\phi) (u\cdot\partial\phi)-\eta^{\mu\nu}{\cal{L}} \  ,
\end{eqnarray}
where $\eta^{\mu\nu}$ denotes the usual Minkowski flat space-time metric tensor. This tensor obeys the condition,  
\begin{eqnarray}
\partial_\mu T^{\mu\nu}=0 \  ,
\end{eqnarray}
however, it is not symmetric. Its antisymmetric part is given by
\begin{eqnarray}
T^{\mu\nu}-T^{\nu\mu}=\lambda\left[u^\mu(\partial^\nu\phi)-u^\nu(\partial^\mu\phi)\right](u\cdot\partial\phi) \  , 
\end{eqnarray}
and this fact is consequence of the Lorentz-breaking term.

An hermitian massive scalar quantum field confined in a 4-dimensional box in thermal equilibrium at temperature $T = 1/ \beta$ can 
be considered as a set of oscillators in thermal equilibrium \cite{merzbacher1998quantum}. This approach is motivated by the 
fact that quanta of the field are being continuously absorbed and re-emitted by the walls, and therefore the temperature of 
the field depends on the temperature of the walls.

Consequently, there is no a pure state but a statistical mixture represented by the density operator
\begin{eqnarray}
 \label{density_operator}
 \hat{\rho} = \frac{e^{-\beta(\hat{H} - \mu \hat{N})}}{Z}.
\end{eqnarray}
The normalization condition gives us
\begin{eqnarray}
 \label{grand_partition_funtion}
 Z = Tr \Big{[}e^{-\beta(\hat{H} - \mu \hat{N})}\Big{]}
\end{eqnarray}
which is called the grand partition function. In the above expressions, $\hat{H}$ is the Hamiltonian operator, $\hat{N}$ is the number operator  and $\mu$ is the chemical potential.

The ensemble average of any physical quantity represented by an operator $\hat{O}$ may be computed by application of the 
formula
\begin{eqnarray}
 \langle \hat{O} \rangle = {\rm Tr} \left(\hat{\rho} \hat{O}\right).
\end{eqnarray}
Applying this relation to the evaluation of the average occupation numbers, $\hat{N_n}$, we obtain
\begin{eqnarray}
 \label{occupation_numbers}
 \langle \hat{N_n} \rangle = \langle \hat{a}^{\dagger}_n \hat{a}_n \rangle = \frac{1}{e^{\beta(\omega_n - \mu)}-1} \  ,
\end{eqnarray}
where $\omega_n$ represents single particle levels of energy. For  a bosonic field we must have $\omega_0 \geq \mu$,  with $\omega_0$ being the minimum of energy. In the high temperature limit the 
expression \eqref{occupation_numbers} can be cast as 
\begin{eqnarray}
 \label{classical_occupation_number}
 \langle \hat{N_n} \rangle = e^{- \beta(\omega_n - \mu)},
\end{eqnarray}
this is, the Maxwell-Boltzmann distribution \cite{merzbacher1998quantum}.

\section{Thermal corrections for Casimir effect}\label{thermal_corrections}
In this section we want to compute the thermal corrections for the Casimir energy admitting that the bosonic quantum field obeys  either Dirichlet, Neumann,  or mixed boundary conditions  on two large parallel plates.  We will assume that the plates are perpendicular to the $z-$axis as shown in Fig. $1$.
\begin{figure}[!htb]
	\label{fig1}
	\centering
	\includegraphics[scale=0.3]{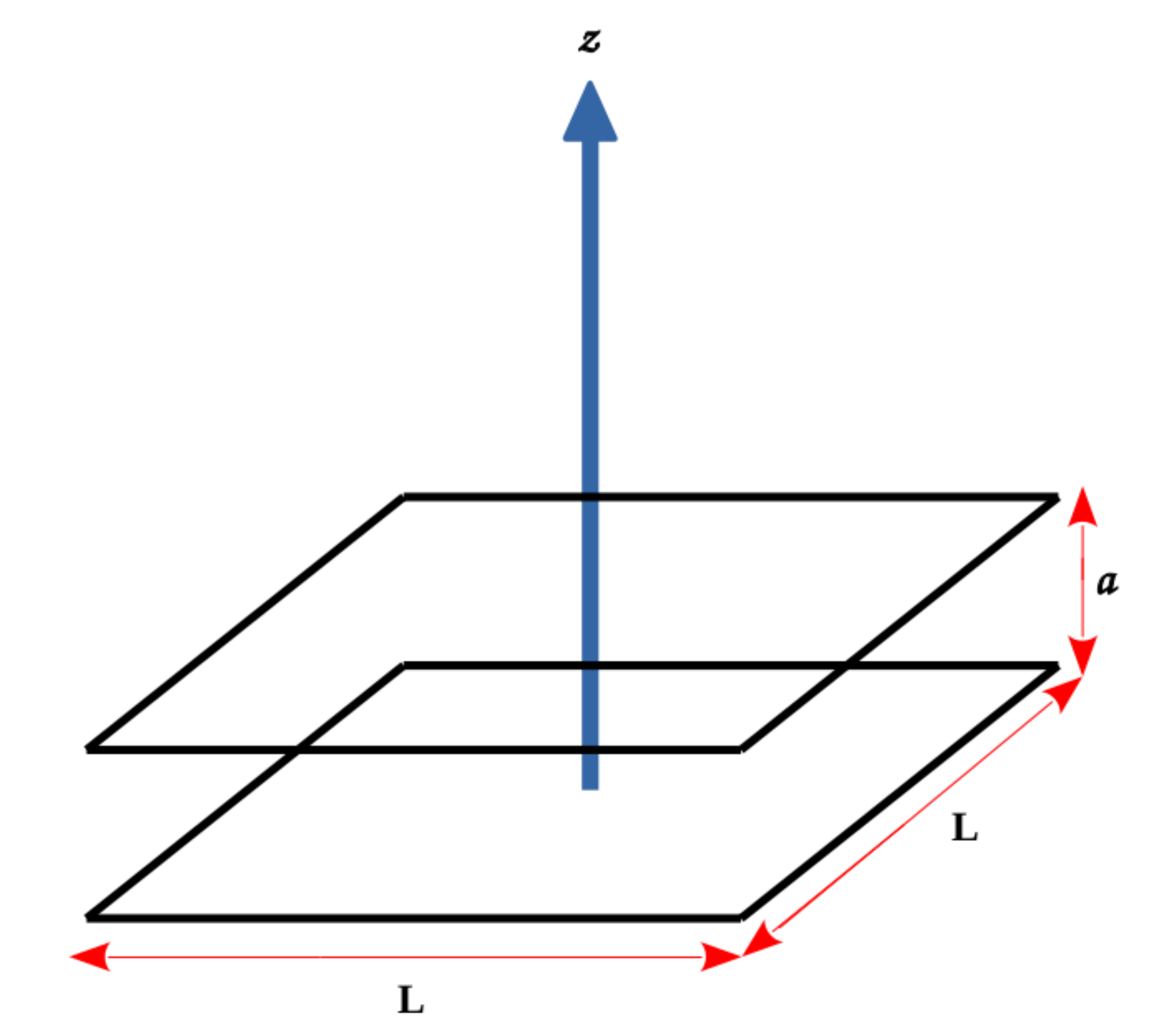}
	\caption{Two parallel plates with area $L^{2}$ separated by a distance $a \ll L$.}
	\label{placas_paralelas}
\end{figure}
\subsection{Dirichlet boundary condition}

In this subsection, we will solve the modified Klein-Gordon equation \eqref{klein_gordon_equation} by imposing the the Dirichlet 
boundary conditions on the two plates, as given below:
\begin{eqnarray}
\label{Dirichlet_condition}
 \phi(x)\Big{|}_{z=0} = \phi(x)\Big{|}_{z=a}.
\end{eqnarray}

Adopting the standard procedure described in textbooks on quantum field theory, one finds the field operator \cite{Cruz:2017kfo}:
\begin{eqnarray}
\label{Dirichlet_operador_campo}
\hat{\phi}(x) = \int d^2\vec{k} \sum_{n=1}^{\infty} \sqrt{ \frac{1}{(2\pi)^2\omega_{\vec{k},n}a} }
\sin \Big{(} \frac{n\pi}{a}z \Big{)} \Big{[} \hat{a}_n(\vec{k})e^{-ikx}+\hat{a}^{\dagger}_n(\vec{k})e^{ikx} \Big{]},
\end{eqnarray}
where
\begin{eqnarray}
 kx=\omega_{\vec{k},n}t-k_{x}x-k_{y}y \  .
\end{eqnarray}
In (\ref{Dirichlet_operador_campo}),  $\hat{a}_{n}(\vec{k})$ and $\hat{a}^{\dagger}_{n}(\vec{k})$ represent the annihilation and creation operators, respectively,  characterized by the set of quantum numbers $\sigma = \{k_{x},k_{y},n\}$.

In what follows we will assume different directions for the Lorentz-breaking constant vector $u^\mu$.

\subsubsection{Time-like vector case}

As our first application we choose the Lorentz-breaking $4-$vector being time-like:
\begin{eqnarray}
 \label{Dirichlet_u_timelike}
 \begin{aligned}
  u^{(0)} = (1,0,0,0)  \  .
 \end{aligned}
\end{eqnarray}
In this case, the Hamiltonian operator $\hat{H}$, can be written as
\begin{eqnarray}
 \label{Dirichlet_Hamiltonian_t}
 \hat{H} = \frac{(1+\lambda)}{2} \int d^2\vec{k} \sum_{n=1}^{\infty} \omega_{\vec{k},n} \Bigg{[} 2\hat{a}^{\dagger}_n(\vec{k})
 \hat{a}_n(\vec{k})+\frac{L^2}{(2\pi)^2} \Bigg{]},
\end{eqnarray}
with the dispersion relation is
\begin{eqnarray}
 \label{Dirichlet_dispersion_t_0}
 \omega_{\vec{k},n} = \sqrt{ \frac{1}{(1+\lambda)} \Big{[} k^2_x + k^2_y + \Big{(} \frac{n\pi}{a} \Big{)}^2 + m^2 \Big{]} }.
\end{eqnarray}

Consequently, the energy is obtained by taking the thermal average of $\hat{H}$:
\begin{eqnarray}
 \label{Dirichlet_energy_t}
 E = \langle \hat{H} \rangle = E_0 + E_T \  ,
\end{eqnarray}
where the first term, $E_0$, is the vacuum energy and the second one, $E_T$, is the 
thermal correction. Since $E_0$ has been analyzed in our previous paper \cite{Cruz:2017kfo}, here we are mainly interested in the calculation of $E_T$.

Therefore, the thermal correction to the energy is given by
\begin{eqnarray}
 \label{Dirichlet_energy_temperature_t_0}
 \begin{aligned}
 E_T &= (1+\lambda) \int d^2\vec{k} \sum_{n=1}^{\infty} \omega_{\vec{k},n} \langle 
 \hat{a}^{\dagger}_n(\vec{k})\hat{a}_n(\vec{k}) \rangle \  .
 \end{aligned}
\end{eqnarray}
By using the relation for the occupation number \eqref{occupation_numbers}, we find
\begin{eqnarray}
 \label{Dirichlet_energy_temperature_t_1}
 \begin{aligned}
 E_T = \frac{(1+\lambda)}{(2\pi)^2}L^2 \int d^2\vec{k} \sum_{n=1}^{\infty} \frac{\omega_{\vec{k},n}}{e^{\beta(\omega_{\vec{k},n} 
 - \mu)}-1} .
 \end{aligned}
\end{eqnarray}
Now, to make the above expression more workable we will use the expansion below,
\begin{eqnarray}
 \label{relation_1}
 \frac{1}{e^z-1} = \sum_{j=1}^{\infty} e^{-jz}  \  .
\end{eqnarray}
So, equation \eqref{Dirichlet_energy_temperature_t_1} become 
\begin{eqnarray}
\label{Dirichlet_energy_temperature_t_2}
E_T = \frac{(1+\lambda)}{(2\pi)^2}L^2 \sum_{j=1}^{\infty} e^{j\beta \mu} \int d^2\vec{k} \sum_{n=1}^{\infty} 
 \omega_{\vec{k},n} e^{-j\beta \omega_{\vec{k},n}}.
\end{eqnarray}

In order to develop the summation on the integer quantum number $n$, we will use the Abel-Plana summation formula \cite{Bordag:2009zzd,Saharian:2007ph}:
\begin{eqnarray}
 \label{Abel_Plana_1}
 \sum_{n=0}^{\infty} F(n)= \frac{1}{2}F(0) + \int_{0}^{\infty}dtF(t) + i\int_{0}^{\infty}\frac{dt}{e^{2\pi t}-1}\Big{[} F(it) - 
 F(-it) \Big{]}.
\end{eqnarray}
Performing in \eqref{Dirichlet_energy_temperature_t_2} a change of coordinates in the Cartesian plane $(k_x,k_y)$ to polar ones, $(k=\sqrt{k^2_x+k^2_y}, \ \theta)$, and integrating over the angular variable, we get
\begin{eqnarray}
 \label{Dirichlet_energy_temperature_t_3}
 E_T&=& \frac{(1+\lambda)}{2\pi}L^2 \sum_{j=1}^{\infty} e^{j\beta \mu} \int_0^{\infty} dkk \Bigg{\{} -\frac{1}{2}F(0) + 
 \int_{0}^{\infty}dtF(t)  \nonumber \\ 
 &+& i\int_{0}^{\infty}\frac{dt}{e^{2\pi t}-1}\Big{[} F(it) - F(-it) \Big{]} \Bigg{\}}  \  ,
\end{eqnarray}
with the function $F(n)$ defined by 
\begin{eqnarray}
 F(n) = \omega_{k,n} e^{-j \beta \omega_{k,n}} \ \ \ \ \ \ \text{being} \ \ \ \ \ \ \omega_{k,n} = \sqrt{ \frac{1}{1+\lambda} 
 \Big{[} k^2 + \Big{(} \frac{n\pi}{a} \Big{)}^2 + m^2 \Big{]}}  \ .
\end{eqnarray}

At this point we would like to call attention to the fact that the first term on the right-hand side of \eqref{Dirichlet_energy_temperature_t_3} represents the energy in the presence  of a single plate, and the second one is connected with energy without boundary. Although both energies are finite, here we will discard them because our main aim is the calculation of the Casimir energy. As a result, the finite-temperature correction for Casimir energy, $E_{T,C}$, is given by
\begin{eqnarray}
 \label{Dirichlet_Casimir_energy_t_0}
 \begin{aligned}
 E_{T,C} &= \frac{1+\lambda}{2\pi} L^2 i \sum_{j=1}^{\infty} e^{j\beta \mu} \int_0^{\infty} dkk \int_{0}^{\infty}\frac{dt}
 {e^{2\pi t}-1}\Big{[} F(it) - F(-it) \Big{]} \\
 &= \frac{\sqrt{1+\lambda}}{2\pi} L^2 i \sum_{j=1}^{\infty} e^{j\beta \mu} \int_0^{\infty} dkk \int_{0}^{\infty}\frac{dt}
 {e^{2\pi t}-1} \Bigg{[} \sqrt{k^2 + m^2 + \Big{(}\frac{i\pi t}{a}\Big{)}^2}  \\
 & \times e^{-\frac{j\beta}{\sqrt{1+\lambda}} \sqrt{k^2 + m^2 + \Big{(}\frac{i\pi t}{a}\Big{)}^2}} - 
 \sqrt{k^2 + m^2 + \Big{(}\frac{-i\pi t}{a}\Big{)}^2} e^{-\frac{j\beta}{\sqrt{1+\lambda}} \sqrt{k^2 + m^2 +
 \Big{(}\frac{-i\pi t}{a}\Big{)}^2}} \Bigg{]}.
 \end{aligned}
\end{eqnarray}
Performing two changes of variables: $u=\pi t/a$ and $x^2=k^2+m^2$, we get
\begin{eqnarray}
 \label{Dirichlet_Casimir_energy_t_1}
 \begin{aligned}
 E_{T,C} &= \frac{\sqrt{1+\lambda}}{2\pi^2} L^2 a i \sum_{j=1}^{\infty} e^{j\beta \mu} \int_m^{\infty} dxx 
 \int_{0}^{\infty}du  \\ & \times \frac{\sqrt{x^2 + (iu)^2} e^{-\frac{j\beta}{\sqrt{(1+\lambda)}} 
 \sqrt{x^2 + (iu)^2}} - \sqrt{x^2 + (-iu)^2} e^{-\frac{j\beta}{\sqrt{(1+\lambda)}} \sqrt{x^2 + (-iu)^2}}}{e^{2au}-1}.
 \end{aligned}
\end{eqnarray}
The integral over the $u$ variable must be considered in two cases: 
\begin{eqnarray}
\label{ident1}
 \sqrt{x^2 + (\pm iu)^2}=\left\{ 
\begin{array}{ll}
\sqrt{x^2 - u^2} \ \ \ \ \ {\rm for} \ \ x>u  \ , & \\
 \pm i \sqrt{u^2 - x^2}\   \ {\rm for} \ \  x<u \ . & \\
\end{array}
\right.
\end{eqnarray}

Thus, we divide the integration interval into two intervals: the first with $x<u$ and that second one with $x>u$. One finds that the integral over $u$ in the interval $[0,x]$ vanishes. So, we get
\begin{eqnarray}
 \label{Dirichlet_Casimir_energy_t_2}
 \begin{aligned}
 E_{T,C} &= - \frac{\sqrt{1+\lambda}}{\pi^2} L^2 a \sum_{j=1}^{\infty} e^{j\beta \mu} \int_m^{\infty} dxx 
 \int_{x}^{\infty}du \frac{\sqrt{u^2 - x^2}}{e^{2au}-1} \cos \Bigg{(} \frac{j \beta}{\sqrt{1+\lambda}}\sqrt{u^2 - x^2} 
 \Bigg{)} \  .
 \end{aligned}
\end{eqnarray}
Performing a convenient changing of variables,
\begin{eqnarray}
 \label{Dirichlet_Casimir_energy_t_4}
 \begin{aligned}
 E_{T,C} &= - \frac{\sqrt{1+\lambda}}{\pi^2} \frac{L^2}{a^3} \sum_{j=1}^{\infty} e^{j\beta \mu} \int_{am}^{\infty} dzz^3 
 \int_{1}^{\infty}dy \frac{\sqrt{y^2 - 1}}{e^{2zy}-1} \cos \Bigg{(} \frac{j \beta}{a \sqrt{1+\lambda}} z \sqrt{y^2 - 1} 
 \Bigg{)}.
 \end{aligned}
\end{eqnarray}

In Appendix \ref{appA}, it is explicitly proved that the above integral can be expressed in term of modified Bessel function of second kind, $K_{\nu}(z)$ \cite{abramowitz:1966}:
\begin{eqnarray}
 \label{Dirichlet_Casimir_energy_t_5}
 \begin{aligned}
 E_{T,C} = - \frac{\sqrt{1+\lambda} L^2 m^2}{\pi^2 a} \sum_{j=1}^{\infty} e^{j\beta \mu} \frac{\partial}{\partial 
 \alpha_j} \Bigg{[} \alpha_j \sum_{n=1}^{\infty} \frac{K_2 \Big{(}am \sqrt{\alpha_j^2+4n^2}\Big{)}}{\alpha_j^2+4n^2} \Bigg{]}  \  ,
 \end{aligned}
\end{eqnarray}
where we have used the notation
\begin{eqnarray}
 \label{Dirichelet_alpha_parammeter_t}
 \begin{aligned}
  \alpha_j = \frac{j \beta}{a \sqrt{1+\lambda}}  \  .
 \end{aligned}
\end{eqnarray}

Unfortunately it is not possible to present a closed expression for the sum over $n$ in \eqref{Dirichlet_Casimir_energy_t_5}. So, in order to provide some informations about the behavior of the $E_{T,C}$, in Appendix \ref{appA} we present the most relevant steps adopted in this present calculation and next ones, in order to obtain approximated results for $E_{T,C}$, considering three different regimes of the dimensionless parameter $M=am$.
\begin{itemize}
 \item For the case $am \gg 1$, we can see that the most relevant term in \eqref{Dirichlet_Casimir_energy_t_5} is given by $n=1$. So using the asymptotic expression for the modified Bessel function for large arguments \cite{abramowitz:1966} and after the derivation with respect to $\alpha_j$, we obtain,
 \begin{eqnarray}
 \label{Dirichlet_Casimir_energy_t_6}
 \begin{aligned}
 \frac{E_{T,C}}{L^2} \approx - \frac{\sqrt{1+\lambda} \sqrt{m^3}}{2 \sqrt{2 \pi^3} \sqrt{a^3}} \sum_{j=1}^{\infty} 
 \Bigg{[} \frac{8-\alpha_j^2\Big{(}3+2 am \sqrt{\alpha_j^2+4}\Big{)}}{(\alpha_j^2+4)^{9/4}} \Bigg{]} 
 e^{-am \sqrt{\alpha_j^2+4}+j\beta \mu} .
 \end{aligned}
\end{eqnarray}
However it is still not possible to obtain a closed result for the summation over $j$. For this reason, we have to take another limit. We will assume that $\beta/a=1/(Ta)>>1$. That means that for fixed distance between the plates, we are considering a low-temperature regime. In fact this limit is the most relevant one. So, the summation in $j$ can be trivially  performed:
 \begin{eqnarray}
 \label{Dirichlet_Casimir_energy_t_7}
 \begin{aligned}
 \frac{E_{T,C}}{L^2} \approx am\sqrt{\frac{1+\lambda}{2}} \left(\frac{\sqrt{1+\lambda}m}{\pi \beta}\right)^{3/2}e^{-\beta\Big{(}\frac{m}{\sqrt{1+ \lambda}}-\mu\Big{)}}.
 \end{aligned}
\end{eqnarray}
We note in this case that the energy per unit area decays exponentially.
 \item For case $am = 0$, i.e., for massless field, $m=0$, we have to take $\mu = 0$. Using the asymptotic expression for the modified Bessel function, $K_\nu(z)$, for small arguments and after  taking the derivative with respect to $\alpha_j$, we obtain a closed expression below,
 \begin{eqnarray}
 \label{Dirichlet_Casimir_energy_t_8}
  \frac{E_{T,C}}{L^2}& =& - \frac{\sqrt{1+\lambda}}{8 \pi ^2 a^3} \sum_{j=1}^{\infty} \frac{1}{\alpha_j^4}  \Bigg{[} 24 - 
  4\pi \alpha_j \text{coth} \Big{(}\frac{\pi \alpha_j}{2}\Big{)} - 2\pi^2\alpha_j^2\text{csch}^2\Big{(}\frac{\pi \alpha_j}{2}
  \Big{)} \nonumber \\ 
  & -& \pi^3\alpha_j^3\text{coth}\Big{(}\frac{\pi \alpha_j}{2}\Big{)}\text{csch}^2\Big{(}\frac{\pi \alpha_j}{2}
  \Big{)} \Bigg{]}  \  .
\end{eqnarray}
Again, there is no closed expression for the summation over $j$ in the above expression. Considering again the low-temperature limit,  $\beta/a \gg 1$, the summation over $j$ is promptly obtained.  We get
\begin{eqnarray}
 \label{Dirichlet_Casimir_energy_t_9}
 \begin{aligned}
  \frac{E_{T,C}}{L^2} \approx - \frac{(1+\lambda)^2}{30 \pi  \beta ^4} \left[\pi ^3 a \sqrt{1+\lambda}-15 \beta  \zeta (3)\right]  \  ,
 \end{aligned}
\end{eqnarray}
where $\zeta(z)$ represents the Riemann zeta function. In fact for the specific case, $\zeta(3)\approx 1.202$.
 \item Finally let us consider $am \ll 1$.  For this case the expression \eqref{Dirichlet_Casimir_energy_t_5} is not convenient. In order to find a more convenient expression to analyze this limit, we will use the identity below \cite{deMello:2012xm}:
 	\begin{eqnarray}
 	\label{sum1}
 	\sum_{n=1}^{\infty} \cos(n \alpha) f_{\nu} (c \sqrt{b^2+a^2 n^2}) = -\frac12f_\nu(cb)+\frac12\frac{\sqrt{2 \pi}}{a c^{2\nu}} \sum_{n=-\infty}^{\infty}
 	w_n^{2\nu-1} f_{\nu-1/2}(bw_n) \  ,
 	\end{eqnarray}
 	with $a,b,c > 0$ and $w_n = \sqrt{(2\pi n + \alpha)^2/a^2 + c^2}$, being the function $f_\nu(z)$ defined as below:
 	\begin{eqnarray}
 	f_\nu(z)=\frac{K_\nu(z)}{z^\nu} \  .
 	\end{eqnarray}
 	Using the above identity, Eq. \eqref{sum1}, taking $\alpha=0$ and adapting the other parameters to our specific problem, it is possible to develop an expansion in power of $ma$ for thermal Casimir energy. Defining a new parameter $\sigma$ as
 	\begin{eqnarray}
 	\label{sigma_parameter}
 	\alpha_j=j\sigma \ , \ {\rm with} \ \ \sigma=\frac{\beta}{a\sqrt{1+\lambda}}
 	\end{eqnarray}
 	and considering the low-temperature limit, we found\footnote{The details of the calculations are explicitly presented in the Appendix \ref{appA}. }:
 \begin{eqnarray}
  \label{Dirichlet_Casimir_energy_t_10}
  \begin{aligned}
   \frac{E_{T,C}}{L^2} \approx - \frac{\sqrt{1+\lambda}}{\pi^2 a^3} \sum_{j=1}^{\infty} &\Bigg{[} \frac{a^3m^3}{2\sigma}
   \frac{K_1(jam\sigma)}{j} + \frac{3a^2m^2}{2\sigma^2} \frac{K_2(jam\sigma)}{j^2} - \\ & - 
   \frac{\pi[2+jam\sigma(2+jam\sigma)]}{4j^3\sigma^3} e^{-jam\sigma} \Bigg{]} e^{j \beta \mu} .
  \end{aligned}
 \end{eqnarray}
 Unfortunately it is impossible to obtain a closed expression for the summation over $j$. So, at this moment we will consider two possible cases: $m\beta > 1$ and 
 $m\beta < 1$.
 \begin{itemize}
  \item For case $m\beta > 1$, the dominant contribution comes from the term with $j=1$, and in this case, we can use the asymptotic expression for the modified Bessel function of second kind for large argument. Doing this we obtain,
  \begin{eqnarray}
   \label{Dirichlet_Casimir_energy_t_11}
   \begin{aligned}
   \frac{E_{T,C}}{L^2} & \approx \frac{(1+\lambda)}{4\pi^{3/2}} \Bigg{[}\frac{\sqrt{\pi}m^2}{\beta}-\frac{\sqrt{2}(1+\lambda)
   ^{1/4}am^{5/2}}{\beta^{3/2}}\Bigg{]}e^{-\beta \Big{(}\frac{m}{\sqrt{1+\lambda}}-\mu \Big{)}}.
   \end{aligned}
  \end{eqnarray}
  This case corresponds to the situation with $m/T>1$. 
  \item For case $m\beta < 1$ we have to take $\mu = 0$. In this case, the situation is much more delicate. The summation over $j$ involving the modified Bessel function can only be evaluated approximately. After some intermediate steps we get,
  \begin{eqnarray}
   \label{Dirichlet_Casimir_energy_t_12}
    \frac{E_{T,C}}{L^2}&\approx& -\frac{(1+\lambda)}{120 \pi \beta ^4}  \Big{[}4 \pi ^3 a (\lambda +1)^{3/2}-60 \beta  (\lambda +1) \zeta (3)\nonumber\\
    &-&5 \beta ^2 m^2 \left(\pi  a \sqrt{\lambda +1}-3 \beta \right) \Big{]} .
  \end{eqnarray}
  We can observe that  the zero mass limit result, \eqref{Dirichlet_Casimir_energy_t_9}, is trivially obtained; moreover in \eqref{Dirichlet_Casimir_energy_t_12} there appears an extra term proportional to $(\beta m)^2$ .  
 \end{itemize}
\end{itemize}

Finally we would like to close this subsection by providing, for the massless case, the leading terms for the total Casimir energy considering the zero temperature result obtained in \cite{Cruz:2017kfo}. It reads,
\begin{equation}
\frac{E_C}{L^2} \approx - \frac{(1+\lambda)^{1/2}\pi^2}{2a^3} \Big{(} \frac{1}{720} - \frac{(1+\lambda)^{3/2}\zeta(3)a^3}{\pi^3\beta^3} \Big{)}.
\end{equation}
By the above expression we can see that the Lorentz violation parameter, $\lambda$, appears modifying both contributions of energies; moreover the thermal correction is of order $O(Ta)^3$.

\subsubsection{Spacelike vector case}

For this case there are three distinct directions for the 4-vector $u^{\mu}$. They are: $u^{(1)} = (0,1,0,0)$, $u^{(2)} = (0,0,1,0)$ and  $u^{(3)} = (0,0,0,1)$. For the two first vectors, the dispersion relations yield the same corrections to the energy. So, 
let us consider
\begin{eqnarray}
 \label{Dirichlet_u_spacelike_x}
 \begin{aligned}
  u^{(1)} = (0,1,0,0).
 \end{aligned}
\end{eqnarray}

The corresponding Hamiltonian operator $\hat{H}$ reads \cite{Cruz:2017kfo},
\begin{eqnarray}
 \label{Dirichlet_Hamiltonian_x}
 \begin{aligned}
  \hat{H} = \frac{1}{2} \int d^2\vec{k} \sum_{n=1}^{\infty} \omega_{\vec{k},n} \Bigg{[} 2\hat{a}_n^{\dagger}(\vec{k})
  \hat{a}_n(\vec{k}) + \frac{L^2}{(2\pi)^2} \Bigg{]},
 \end{aligned}
\end{eqnarray}
with the dispersion relation now being
\begin{eqnarray}
 \label{Dirichlet_dispersion_relation_x}
 \begin{aligned}
  \omega_{\vec{k},n} = \sqrt{(1-\lambda)k_x^2 + k_y^2 + \Big{(} \frac{n\pi}{a} \Big{)}^2 +m^2}.
 \end{aligned}
\end{eqnarray}
Consequently, the correction to energy generated by finite temperature is given by
\begin{eqnarray}
 \label{Dirichlet_energy_temperature_x_0}
 \begin{aligned}
  E_T = \int d^2\vec{k} \sum_{n=1}^{\infty} \omega_{\vec{k},n} \langle \hat{a}_n^{\dagger}(\vec{k})\hat{a}_n(\vec{k}) \rangle .
 \end{aligned}
\end{eqnarray}

Using the thermal average relation for the occupation number \eqref{occupation_numbers},  and the expansion \eqref{relation_1},  we get
\begin{eqnarray}
 \label{Dirichlet_energy_temperature_x_1}
 \begin{aligned}
  E_T &= \frac{L^2}{(2\pi)^2} \sum_{j=1}^{\infty} e^{j\beta \mu} \int d^2\vec{k} \sum_{n=1}^{\infty} 
  \omega_{\vec{k},n}e^{-j \beta \omega_{\vec{k},n}}.
 \end{aligned}
\end{eqnarray}

Performing in the above equation a change coordinates from Cartesian to polar system, and still using the summation formula  \eqref{Abel_Plana_1} we get
\begin{eqnarray}
 \label{Dirichlet_energy_temperature_x_2}
  E_T &=& \frac{1}{2\pi \sqrt{1-\lambda}} L^2 \sum_{j=1}^{\infty} e^{j\beta \mu} \int_0^{\infty} dkk \Bigg{\{} -\frac{1}{2}F(0)
  +\int_0^{\infty}dtF(t) \nonumber\\
  && + i \int_0^{\infty} \frac{dt}{e^{2\pi t}-1} \Big{[} F(it) - F(-it) \Big{]} \Bigg{\}} \ ,
\end{eqnarray}
with the function $F(n)$ given by
\begin{eqnarray}
 \begin{aligned}
 F(n) = \omega_{k,n} e^{-j\beta \omega_{k,n}} \ \ \ \ \text{and} \ \ \ \  \omega_{k,n} = \sqrt{k^2 + m^2 + \Big{(} 
 \frac{n\pi}{a} \Big{)}^2}.
 \end{aligned}
\end{eqnarray}
Consequently we have the following thermal Casimir energy
\begin{eqnarray}
 \label{Dirichlet_Casimir_energy_x_0}
 E_{T,C} &=& \frac{1}{2\pi \sqrt{1-\lambda}} L^2 i \sum_{j=1}^{\infty} e^{j\beta \mu} \int_0^{\infty} dkk \int_0^{\infty} 
 \frac{dt}{e^{2\pi t}-1} \Big{[} F(it) - F(-it) \Big{]} \nonumber\\
 &=&\frac{1}{2\pi \sqrt{1-\lambda}} L^2 i \sum_{j=1}^{\infty} e^{j\beta \mu} \int_0^{\infty} dkk \int_0^{\infty} 
 \frac{dt}{e^{2\pi t}-1} \Bigg{[} \sqrt{k^2 + m^2 + \Big{(}\frac{it\pi}{a}\Big{)}^2} \nonumber\\\
 &&\times e^{-j\beta \sqrt{k^2 + m^2 + \Big{(}\frac{it\pi}{a}\Big{)}^2}} - \sqrt{k^2 + m^2 + \Big{(}\frac{-it\pi}{a}\Big{)}^2}
 e^{-j\beta \sqrt{k^2 + m^2 + \Big{(}\frac{-it\pi}{a}\Big{)}^2}} \Bigg{]} \ .
\end{eqnarray}

Performing the same two changes of variables as before, $u = \frac{\pi t}{a}$ and $x^2 = k^2 + m^2$, we get
\begin{eqnarray}
 \label{Dirichlet_Casimir_energy_x_1}
 \begin{aligned}
 E_{T,C} &= \frac{1}{2\pi^2 \sqrt{1-\lambda}} L^2 a i \sum_{j=1}^{\infty} e^{j\beta \mu} \int_m^{\infty} dxx \int_0^{\infty} 
 du \\
 & \times \frac{\sqrt{x^2 + (iu)^2} e^{-j\beta \sqrt{x^2 + (iu)^2}} - \sqrt{x^2 + (-iu)^2} e^{-j\beta \sqrt{x^2 + (-iu)^2}}}
 {e^{2au}-1}.
 \end{aligned}
\end{eqnarray}

Dividing the interval of integration over $u$ in two parts, that is, $[0,x]$ and $[x,\infty)$, and considering the identity \eqref{ident1}, we get
\begin{eqnarray}
 \label{Dirichlet_Casimir_energy_x_2}
 \begin{aligned}
 E_{T,C} = - \frac{1}{\pi^2 \sqrt{1-\lambda}} L^2 a \sum_{j=1}^{\infty} e^{j\beta \mu} \int_m^{\infty} dxx 
 \int_x^{\infty} du \frac{\sqrt{u^2 - x^2}}{e^{2au}-1} \cos \Big{(} j \beta \sqrt{u^2 - x^2} \Big{)} \  .
 \end{aligned}
\end{eqnarray}
Once more, by performing another changing of variables,  $u = xy$ and after $z = ax$, we arrive at
\begin{eqnarray}
 \label{Dirichlet_Casimir_energy_x_3}
 \begin{aligned}
 E_{T,C} = - \frac{L^2}{\pi^2 \sqrt{1-\lambda} a^3} \sum_{j=1}^{\infty} e^{j\beta \mu} \int_{am}^{\infty} dzz^3 
 \int_1^{\infty} dy \frac{\sqrt{y^2 - 1}}{e^{2zy}-1} \cos \Bigg{(} \frac{j \beta}{a}z \sqrt{y^2 - 1} \Bigg{)}.
 \end{aligned}
\end{eqnarray}
Again, the above integral can be expressed in terms of the modified Bessel function, $K_{\nu}(z)$, as it is shown in Appendix \ref{appA}:
\begin{eqnarray}
 \label{Dirichlet_Casimir_energy_x_4}
 \begin{aligned}
 E_{T,C} = -\frac{L^2 m^2}{\pi ^2 \sqrt{1-\lambda } a} \sum_{j=1}^{\infty} e^{j \beta \mu} \frac{\partial}{\partial 
 \alpha_j} \Bigg{[} \alpha_j \sum_{n=1}^{\infty} \frac{K_2\Big{(} am \sqrt{\alpha_j^2+4n^2}\Big{)}}{\alpha_j^2+4n^2} \Bigg{]}  \  ,
 \end{aligned}
\end{eqnarray}
with parameter $\alpha_j$ defined by
\begin{eqnarray}
 \begin{aligned}
  \alpha_j = \frac{j \beta}{a}  \  .
 \end{aligned}
\end{eqnarray}
Here, in this analysis, we develop a similar procedures as adopted in previous analysis. So, we will skip parts of the explicit calculations.

 To develop the sum over $n$, we take the asymptotic limits $am \gg 1$, $am = 0$ and  $am \ll 1$.

\begin{itemize}
 \item For the case $am \gg 1$ the most relevant term is given by $n=1$. After some intermediate steps already mentioned  earlier, we obtain:
 \begin{eqnarray}
 \label{Dirichlet_Casimir_energy_x_5}
 \begin{aligned}
 \frac{E_{T,C}}{L^2} \approx - \frac{m^{3/2}}{\sqrt{8 \pi^3} \sqrt{1-\lambda} a^{3/2}} \sum_{j=1}^{\infty} \Bigg{[}
 \frac{8-\alpha_j^2\Big{(}3+2am\sqrt{\alpha_j^2+4}\Big{)}}{(\alpha_j^2+4)^{9/4}}\Bigg{]} e^{-am\sqrt{\alpha_j^2+4}+j\beta \mu} .
 \end{aligned}
\end{eqnarray}
To make the summation in $j$, we consider the low-temperature limit, i.e., $\beta/a \gg 1$. So we get, 
\begin{eqnarray}
 \label{Dirichlet_Casimir_energy_x_6}
 \begin{aligned}
 \frac{E_{T,C}}{L^2} \approx \frac{a m^{5/2}}{\sqrt{2} \pi^{3/2} \sqrt{1-\lambda } \beta^{3/2}} e^{-\beta(m-\mu)}.
 \end{aligned}
\end{eqnarray}
Clearly, this contribution presents an exponential decay.
 \item For the case $am = 0$, we have to take $\mu = 0$. We find,
 \begin{eqnarray}
 \label{Dirichlet_Casimir_energy_x_7}
 \begin{aligned}
  \frac{E_{T,C}}{L^2} = - \frac{1}{8\pi^2 \sqrt{1-\lambda} a^3} \sum_{j=1}^{\infty} \frac{1}{\alpha_j^4} & \Bigg{[} 24 - 
  4\pi \alpha_j \text{coth}\Big{(}\frac{\pi \alpha_j}{2}\Big{)} - 2\pi^2 \alpha_j^2\text{csch}^2\Big{(}\frac{\pi \alpha_j}{2}
  \Big{)} - \\ & - \pi^3 \alpha_j^3 \text{coth}\Big{(}\frac{\pi \alpha_j}{2}\Big{)} \text{csch}^2\Big{(}\frac{\pi \alpha_j}{2}
  \Big{)} \Bigg{]}.
 \end{aligned}
\end{eqnarray}
Considering the low-temperature limit it is possible to sum over $j$ and we get
\begin{eqnarray}
 \label{Dirichlet_Casimir_energy_x_8}
 \begin{aligned}
  \frac{E_{T,C}}{L^2} \approx - \frac{1}{30 \pi \sqrt{1-\lambda} \beta ^4} \Big{[}\pi ^3 a-15 \beta \zeta (3)\Big{]} \  .
 \end{aligned}
\end{eqnarray}
 \item For the case $am \ll 1$, we have to adopt another representation to calculate the Casimir energy, as we did  above.   Proceeding in this way, we consider the identity \eqref{sum1}, and defining a new parameter $\sigma$, as 
 \begin{eqnarray}
 \label{sigma_parameter_x}
 \begin{aligned}
 \alpha_j = j \sigma, \ \ \ \ \sigma = \frac{\beta}{a}
 \end{aligned}
 \end{eqnarray}
and assuming the low-temperature limit, we obtain,
 \begin{eqnarray}
  \label{Dirichlet_Casimir_energy_x_9}
   \frac{E_{T,C}}{L^2} &\approx& - \frac{1}{\pi^2 \sqrt{1-\lambda}a^3} \sum_{j=1}^{\infty}  \Bigg{[} \frac{a^3m^3}{2\sigma}
   \frac{K_1(jam\sigma)}{j} + \frac{3a^2m^2}{2\sigma^2} \frac{K_2(jam\sigma)}{j^2} \nonumber\\
   & -&   \frac{\pi[2+jam\sigma(2+jam\sigma)]e^{-jam\sigma}}{4j^3 \sigma^3} \Bigg{]} e^{j\beta \mu} .
 \end{eqnarray}
 Because there is no possibility to express the sum over $j$ in a closed form,  we must consider two possible sub-cases: $m\beta > 1$ and 
 $m\beta < 1$.
 \begin{itemize}
  \item For case $m\beta > 1$, we get
  \begin{eqnarray}
   \label{Dirichlet_Casimir_energy_x_11}
   \begin{aligned}
   \frac{E_{T,C}}{L^2} & \approx \frac{1}{4 \pi ^2 \sqrt{1-\lambda }} \Bigg{[} \frac{\pi m^2}{\beta} - \frac{\sqrt{2 \pi} a 
   m^{5/2}}{\beta^{3/2}} \Bigg{]} e^{-\beta(m-\mu)}.
   \end{aligned}
  \end{eqnarray}
  \item For the case $m\beta < 1$ we have take $\mu = 0$. So after some intermediate steps we get,
  \begin{eqnarray}
   \label{Dirichlet_Casimir_energy_x_12}
   \begin{aligned}
    \frac{E_{T,C}}{L^2} \approx -\frac{1}{120 \pi \sqrt{1-\lambda } \beta ^4} \Big{[}4 \pi ^3 a-60 \beta  \zeta(3) -5\beta^2m^2(\pi a-3\beta)\Big{]}.
   \end{aligned}
  \end{eqnarray}
  Here also, we can see that by the above result, the zero mass limit, \eqref{Dirichlet_Casimir_energy_x_8}, is automatically obtained.
 \end{itemize}
\end{itemize}

For this specific situation of space-like constant vector, the leading contribution to the total Casimir energy for the massless field, can be provided by using the corresponding result of \cite{Cruz:2017kfo}. It reads,
\begin{equation}
\frac{E_C}{L^2} \approx - \frac{(1-\lambda)^{-1/2}\pi^2}{2a^3} \Big{(}\frac{1}{720}-\frac{\zeta(3)a^3}{\pi^3\beta^3} \Big{)}.
\end{equation}
By the above expression we can see that the Lorentz violation parameter, $\lambda$, appears modifying the total energy only.

Finally, let us consider the constant 4-vector orthogonal to the plates:
\begin{eqnarray}
 \begin{aligned}
 u^{(3)} = (0,0,0,1).
 \end{aligned}
\end{eqnarray}
In this case, the  Hamiltonian operator reads,
\begin{eqnarray}
 \label{Dirichlet_Hamiltonian_z}
 \begin{aligned}
  \hat{H} = \frac{1}{2} \int d^2\vec{k} \sum_{n=1}^{\infty} \omega_{\vec{k},n} \Bigg{[} 2\hat{a}_n^{\dagger}(\vec{k})
  \hat{a}_n(\vec{k}) + \frac{L^2}{(2\pi)^2} \Bigg{]} \   ,
 \end{aligned}
\end{eqnarray}
where  the dispersion relation is now given by 
\begin{eqnarray}
 \label{Dirichlet_dispersion_relation_z}
 \begin{aligned}
  \omega_{\vec{k},n} = \sqrt{k_x^2 + k_y^2 + (1-\lambda) \Big{(} \frac{n\pi}{a} \Big{)}^2 + m^2}.
 \end{aligned}
\end{eqnarray}
Consequently, one finds that the finite temperature correction to the Casimir energy is given by
\begin{eqnarray}
 \label{Dirichlet_energy_temperature_z_0}
 \begin{aligned}
  E_T = \int d^2\vec{k} \sum_{n=1}^{\infty} \omega_{\vec{k},n} \langle \hat{a}_n^{\dagger}(\vec{k})\hat{a}_n(\vec{k}) \rangle.
 \end{aligned}
\end{eqnarray}
Again using the relation for the occupation number \eqref{occupation_numbers} together with \eqref{relation_1}, we find
\begin{eqnarray}
 \label{Dirichlet_energy_temperature_z_1}
 \begin{aligned}
  E_T = \frac{L^2}{(2\pi)^2} \sum_{j=1}^{\infty} e^{j\beta \mu} \int d^2\vec{k} \sum_{n=1}^{\infty} \omega_{\vec{k},n}
  e^{-j \beta \omega_{\vec{k},n}}.
 \end{aligned}
\end{eqnarray}

Developing again the summation over the $n$ by using \eqref{Abel_Plana_1} and performing a change of coordinates in the 
plan $(k_x, k_y)$ to polar ones, with
\begin{eqnarray}
 F(n) = \omega_{k,n} e^{-j \beta \omega_{k,n}} \ \ \ \ \text{where} \ \ \ \ \omega_{k,n} = \sqrt{k^2 + m^2 + (1-\lambda)
 \Big{(} \frac{n\pi}{a} \Big{)}^2},
\end{eqnarray}
one finds that the energy is
\begin{eqnarray}
 \label{Dirichlet_energy_temperature_z_2}
  E_T& =& \frac{L^2}{2\pi} \sum_{j=1}^{\infty} e^{j\beta \mu} \int_0^{\infty} dkk \Bigg{\{} -\frac{1}{2}F(0) +
  \int_0^{\infty} dtF(t) \nonumber\\
  &+& i \int_0^{\infty} \frac{dt}{e^{2\pi t}-1} \Big{[} F(it) - F(-it) \Big{]} \Bigg{\}}.
\end{eqnarray}
From the above expression we can conclude that the finite-temperature correction to the Casimir energy is given by
\begin{eqnarray}
 \label{Dirichlet_Casimir_energy_temperature_z_0}
  E_{T,C} &= &\frac{L^2}{2\pi} i \sum_{j=1}^{\infty} e^{j\beta \mu} \int_0^{\infty} dkk \int_0^{\infty} 
  \frac{dt}{e^{2\pi t}-1} \Big{[} F(it) - F(-it) \Big{]} \\
  &=& \frac{L^2}{2\pi} i \sum_{j=1}^{\infty} e^{j\beta \mu} \int_0^{\infty} dkk \int_0^{\infty} \frac{dt}{e^{2\pi t}-1} 
  \Bigg{[} \sqrt{k^2 + m^2 + (1-\lambda) \Big{(} \frac{it\pi}{a} \Big{)}^2} \nonumber\\
  & \times& e^{-j\beta \sqrt{k^2 + m^2 + (1-\lambda)
  \Big{(} \frac{it\pi}{a} \Big{)}^2}} - \sqrt{k^2 + m^2 + (1-\lambda) \Big{(} - \frac{it\pi}{a} \Big{)}^2}\nonumber\\
&\times& e^{-j\beta \sqrt{k^2 + m^2 + (1-\lambda) \Big{(} - \frac{it\pi}{a} \Big{)}^2}} \Bigg{]} \  .
\end{eqnarray}
 Performing the following change of variables $u=\frac{\pi t}{b}$ with $\frac{1}{b}=\frac{\sqrt{(1-\lambda)}}{a}$, and $x^2=k^2+m^2$, we get
\begin{eqnarray}
 \label{Dirichlet_Casimir_energy_temperature_z_1}
  E_{T,C} &= &\frac{L^2 b}{2\pi^2} i \sum_{j=1}^{\infty} e^{j\beta \mu} \int_m^{\infty} dxx \int_0^{\infty} du \nonumber\\
  &\times& \frac{\sqrt{x^2 + (iu)^2}e^{-j\beta \sqrt{x^2 + (iu)^2}} - \sqrt{x^2 + (-iu)^2}e^{-j\beta \sqrt{x^2 + (-iu)^2}}}
  {e^{2bu}-1}  \  .
\end{eqnarray}
Carrying out integral in $u$ in the same way as above, we get
\begin{eqnarray}
 \label{Dirichlet_Casimir_energy_temperature_z_2}
 \begin{aligned}
  E_{T,C} = - \frac{L^2 b}{\pi^2} \sum_{j=1}^{\infty} e^{j\beta \mu} \int_m^{\infty} dxx \int_x^{\infty} du
  \frac{\sqrt{u^2 - x^2}}{e^{2bu}-1} \cos \Big{(} j\beta \sqrt{u^2 - x^2} \Big{)}  \  .
 \end{aligned}
\end{eqnarray}

Finally, performing two changes of variables $u=xy$ and $z=bx$ we find
\begin{eqnarray}
 \label{Dirichlet_Casimir_energy_temperature_z_3}
 \begin{aligned}
  E_{T,C} = - \frac{L^2}{\pi^2 b^3} \sum_{j=1}^{\infty} e^{j\beta \mu} \int_{bm}^{\infty} dzz^3 \int_1^{\infty} dy
  \frac{\sqrt{y^2 - 1}}{e^{2zy}-1} \cos \Big{(} \frac{j\beta}{b} z \sqrt{y^2 - 1} \Big{)}.
 \end{aligned}
\end{eqnarray}
Expressing the integral above in terms of modified Bessel function, $K_{\nu}(z)$, we obtain
\begin{eqnarray}
 \label{Dirichlet_Casimir_energy_temperature_z_4}
 \begin{aligned}
  E_{T,C} = - \frac{L^2 m^2}{\pi^2 b} \sum_{j=1}^{\infty} e^{j \beta \mu} \frac{\partial}{\partial \alpha_j} \Bigg{[} 
  \alpha_j \sum_{n=1}^{\infty} \frac{K_2 \Big{(} bm \sqrt{\alpha_j^2+4n^2} \Big{)}}{\alpha_j^2+4n^2} \Bigg{]},
 \end{aligned}
\end{eqnarray}
where now the parameter $\alpha_j$ is given by
\begin{eqnarray}
 \begin{aligned}
  \alpha_j = \frac{j \beta}{b}.
 \end{aligned}
\end{eqnarray}
Here, again we consider the asymptotic limits: $am \gg 1$, $am = 0$ and $am \ll 1$. 
\begin{itemize}
 \item For case $am \gg 1$ the most relevant term is given by $n=1$:
 \begin{eqnarray}
 \label{Dirichlet_Casimir_energy_z_5}
 \begin{aligned}
 \frac{E_{T,C}}{L^2} \approx -\frac{m^{3/2}}{2 \sqrt{2} \pi ^{3/2} b^{3/2}} \sum_{j=1}^{\infty} \Bigg{[} 
 \frac{8-\alpha_j^2\Big{(}3+2bm\sqrt{\alpha_j^2+4}\Big{)}}{(\alpha_j^2+4)^{9/4}} \Bigg{]} e^{-bm\sqrt{\alpha_j^2+4}+j\beta \mu} \  .
 \end{aligned}
\end{eqnarray}
Considering the low-temperature limit, $\beta/a \gg 1$, we get 
\begin{eqnarray}
 \label{Dirichlet_Casimir_energy_z_6}
 \begin{aligned}
 \frac{E_{T,C}}{L^2} \approx \frac{a m^{5/2}}{\sqrt{2} \pi^{3/2} \sqrt{1-\lambda} \beta^{3/2}} e^{- \beta(m - \mu)}.
 \end{aligned}
\end{eqnarray}
It is observed in this case that the energy per unit area decays exponentially with $\beta m$.
 \item For case $am = 0$ we have to take $\mu = 0$:
 \begin{eqnarray}
 \label{Dirichlet_Casimir_energy_z_7}
 \begin{aligned}
  \frac{E_{T,C}}{L^2} = -\frac{1}{8 \pi ^2 b^3} \sum_{j=1}^{\infty} \frac{1}{\alpha_j^4} & \Bigg{[} 24 - 4\pi \alpha_j
  \text{coth}\Big{(}\frac{\pi \alpha_j}{2}\Big{)} - 2\pi^2 \alpha_j^2\text{csch}^2\Big{(}\frac{\pi \alpha_j}{2}\Big{)} - \\
  & - \pi^3\alpha_j^3\text{coth}\Big{(}\frac{\pi \alpha_j}{2}\Big{)}\text{csch}^2\Big{(}\frac{\pi \alpha_j}{2}\Big{)} \Bigg{]}.
 \end{aligned}
\end{eqnarray}
Considering the low-temperature limit and using the definition \eqref{sigma_parameter_z}, we get
\begin{eqnarray}
 \label{Dirichlet_Casimir_energy_z_8}
 \begin{aligned}
  \frac{E_{T,C}}{L^2} \approx -\frac{1}{30 \pi  \beta ^4} \Bigg{[}\frac{\pi ^3a}{\sqrt{1-\lambda}}-15 \beta  \zeta(3)\Bigg{]} .
 \end{aligned}
\end{eqnarray}
 \item For case $am \ll 1$, we have to use a more convenient expression to compute the Casimir energy, Eq. \eqref{sum1}. Considering the definition
 \begin{eqnarray}
 \label{sigma_parameter_z}
 \begin{aligned}
 \alpha_j = j \sigma, \ \ \ \ \sigma = \frac{\beta}{b} \ \ \ \ \text{with} \ \ \ \ b = \frac{a}{\sqrt{1-\lambda}}  \  ,
 \end{aligned}
 \end{eqnarray}
 and the low-temperature limit, we obtain
 \begin{eqnarray}
  \label{Dirichlet_Casimir_energy_z_9}
   \frac{E_{T,C}}{L^2} &\approx& - \frac{1}{\pi^2 b^3} \sum_{j=1}^{\infty} \Bigg{[} \frac{b^3m^3}{2\sigma} \frac{K_1(jbm\sigma)}
   {j} + \frac{3b^2m^2}{2\sigma^2} \frac{K_2(jbm\sigma)}{j^2} \nonumber\\
   &-& \frac{\pi[2+jbm\sigma(2+jbm\sigma)]e^{-jbm\sigma}}
   {4j^3\sigma^3} \Bigg{]} e^{j\beta \mu}  \  .
 \end{eqnarray}
 As it is impossible to present the sum over $j$ in a closed form, we must consider two  possible  cases: $m\beta > 1$ and 
 $m\beta < 1$.
 \begin{itemize}
  \item For case $m\beta > 1$, we get
  \begin{eqnarray}
   \label{Dirichlet_Casimir_energy_z_10}
   \begin{aligned}
   \frac{E_{T,C}}{L^2} & \approx \frac{1}{4\pi^{3/2}} \Bigg{[} \frac{\sqrt{\pi }m^2}{\beta} -
   am\sqrt{\frac{2}{1-\lambda}} \left(\frac m\beta\right)^{3/2} 
   \Bigg{]} e^{-\beta(m-\mu)} .
   \end{aligned}
  \end{eqnarray}
 \end{itemize}
 \begin{itemize}
  \item For case $m\beta < 1$ we have take $\mu = 0$, we get
  \begin{eqnarray}
   \label{Dirichlet_Casimir_energy_z_11}
    \frac{E_{T,C}}{L^2} &\approx& -\frac{1}{120 \pi \sqrt{1-\lambda} \beta^4}  \Big{[}4 \pi^3 a-60 \beta \sqrt{1-\lambda } \zeta(3) \nonumber\\ 
    & -&5\beta^2m^2(\pi a-3\beta\sqrt{1-\lambda}) \Big{]}  \ .
  \end{eqnarray}
 \end{itemize}
\end{itemize}

As in the previous two subsections, below we provide the leading contribution to the total Casimir energy for the massless field. It reads,
\begin{equation}
\frac{E_C}{L^2} \approx - \frac{(1-\lambda)^{3/2}\pi^2}{2a^3} \Big{(}\frac{1}{720} - \frac{(1-\lambda)^{-3/2}\zeta(3)a^3}{\pi^3\beta^3} \Big{)}.
\end{equation}

\subsection{Neumann boundary condition}

In this section we will investigate the thermal correction to the Casimir energy, admitting that the scalar field obeys the Neumann boundary condition on the plates, as shown below,
\begin{eqnarray}
 \label{Neumann_condition}
 \begin{aligned}
  \frac{\partial \phi(x)}{\partial z} \Bigg{|}_{z=0} = \frac{\partial \phi(x)}{\partial z} \Bigg{|}_{z=a} = 0.
 \end{aligned}
\end{eqnarray}

The solution of the modified Klein-Gordon equation, \eqref{klein_gordon_equation}, compatible with the above boundary condition is \cite{Cruz:2017kfo},
\begin{eqnarray}
 \label{Neumann_operator_field}
 \begin{aligned}
  \hat{\phi}(x) = \int d^2\vec{k} \sum_{n=0}^{\infty} c_n \cos \Big{(} \frac{n\pi}{a} z \Big{)} \Big{[} \hat{a}_n(\vec{k})
  e^{-ikx} + \hat{a}_n^{\dagger}(\vec{k}) e^{ikx} \Big{]},
 \end{aligned}
\end{eqnarray}
where the normalization constant is
\begin{eqnarray}
\label{normalization_constant}
 c_{n}=\left \{\begin{array}{c}
 \sqrt{\frac{1}{2\sqrt{2\pi}\omega_{n}(\vec{k})a}} \ \ \text{for} \ n=0, \\
 \sqrt{\frac{1}{\sqrt{2\pi}\omega_{n}(\vec{k})a}} \ \ \text{for} \text{ } n\geq 0.
\end{array} \right.
\end{eqnarray}

In this case, although we can notice that the field operator is different from the corresponding one obtained by imposing Dirichlet boundary conditions on the fields, the Hamiltonian operator and the dispersion relations  remain the same as  for the Dirichlet boundary condition, for each choice of the 4-vector $u^{\mu}$. So, we will not repeat all the calculations because they are exactly the same. 

\subsection{Mixed boundary condition}

Now, let us consider the situation where the scalar field obeys a Dirichlet boundary condition on one plate and a Neumann boundary condition 
on the other one. In this case, two different configurations take place:
 \begin{eqnarray}
  \phi(\vec{x}) \Big{|}_{z=0} = \frac{\partial \phi(\vec{x})}{\partial z} \Bigg{|}_{z=a} = 0.
 \end{eqnarray}
 \begin{eqnarray}
  \frac{\partial \phi(\vec{x})}{\partial z} \Bigg{|}_{z=0} = \phi(\vec{x}) \Big{|}_{z=a} = 0.
 \end{eqnarray}

After solving the Klein-Gordon equation \eqref{klein_gordon_equation} with these conditions, the field operators read \cite{Cruz:2017kfo}

\begin{eqnarray}
 \label{Misto_campo_operator_a}
 \hat{\phi}_a(x) = \int d^2\vec{k} \sum_{n=0}^{\infty} \sqrt{\frac{1}{(2\pi)^2\omega_{\vec{k},n}a}} \sin \Big{[} \Big{(}
 n+\frac{1}{2} \Big{)}\frac{\pi}{a}z \Big{]} \Big{[} \hat{a}_n(\vec{k})e^{-ikx} + \hat{a}_n^{\dagger}(\vec{k})e^{ikx} \Big{]}
\end{eqnarray}
for the first configuration and
\begin{eqnarray}
 \label{Misto_campo_operator_b}
 \hat{\phi}_b(x) = \int d^2\vec{k} \sum_{n=0}^{\infty} \sqrt{\frac{1}{(2\pi)^2\omega_{\vec{k},n}a}} \cos \Big{[} \Big{(}
 n+\frac{1}{2} \Big{)}\frac{\pi}{a}z \Big{]} \Big{[} \hat{a}_n(\vec{k})e^{-ikx} + \hat{a}_n^{\dagger}(\vec{k})e^{ikx} \Big{]}
\end{eqnarray}
for the second configuration.

Both field operators, $\hat{\phi}_a(x)$ and $\hat{\phi}_a(x)$, provide the same Hamiltonian operator and present the same  dispersion relations, for each cases of constant 4-vector $u^\mu$.

\subsubsection{Time-like vector case}

Let us start our analysis taking a time-like 4-vector, $u^{(0)} = (1,0,0,0)$. In this case  the Hamiltonian operator reads,
\begin{eqnarray}
 \label{Mista_Hamiltonian_operator_t}
 \begin{aligned}
  \hat{H} = \frac{(1+\lambda)}{2} \int d^2\vec{k} \sum_{n=0}^{\infty} \omega_{\vec{k},n} \Bigg{[} 2\hat{a}_n^{\dagger}(\vec{k})
  \hat{a}_n(\vec{k}) + \frac{L^2}{(2\pi)^2} \Bigg{]}  \  ,
 \end{aligned}
\end{eqnarray}
where $\omega_{\vec{k},n}$ satisfies the dispersion relation,
\begin{eqnarray}
 \label{Mista_dispersion_t_0}
 \begin{aligned}
  \omega_{\vec{k},n} = \sqrt{\frac{1}{(1+\lambda)} \Big{[}k_x^2+k_y^2+ \Big{(} \Big{(} n+\frac{1}{2} \Big{)}\frac{\pi}{a}
  \Big{)}^2 +m^2 \Big{]}}.
 \end{aligned}
\end{eqnarray}

The energy of the scalar field is expressed as
\begin{eqnarray}
 \label{Mista_energy_t}
 \begin{aligned}
  E = \langle \hat{H} \rangle = E_0 + E_T  \   .
 \end{aligned}
\end{eqnarray}
We would like to remember that the first term of  right-hand side is the vacuum contribution, while that the second term  is the contribution of finite temperature, given by 
\begin{eqnarray}
 \label{Mista_energy_temperatura_t_0}
 \begin{aligned}
 E_T = (1+\lambda) \int d^2\vec{k} \sum_{n=0}^{\infty} \omega_{\vec{k},n} \langle \hat{a}_n^{\dagger}(\vec{k})
 \hat{a}_n(\vec{k}) \rangle  \  .
 \end{aligned}
\end{eqnarray}
Using the relation for the occupation number \eqref{occupation_numbers}, we get
\begin{eqnarray}
 \label{Mista_energy_temperatura_t_1}
 \begin{aligned}
 E_T =  \frac{(1+\lambda)L^2}{(2\pi)^2} \int d^2\vec{k} \sum_{n=0}^{\infty} \frac{{\omega_{\vec{k},n}}}{e^{\beta(\omega_{\vec{k},n}
 -\mu)}-1}  \  ,
 \end{aligned}
\end{eqnarray}
and still using the relation \eqref{relation_1}, we get 
\begin{eqnarray}
 \label{Mista_energy_temperatura_t_2}
 \begin{aligned}
 E_T = \frac{(1+\lambda)L^2}{(2\pi)^2} \sum_{j=1}^{\infty} e^{j\beta \mu} \int d^2\vec{k} \sum_{n=0}^{\infty} 
 \omega_{\vec{k},n} e^{-j\beta \omega_{\vec{k},n}}.
 \end{aligned}
\end{eqnarray}

Changing the Cartesian coordinates $(k_x,k_y)$ to polar ones, and using the Abel-Plana summation formula for half-integer numbers \cite{Bordag:2009zzd, Saharian:2007ph},
\begin{eqnarray}
 \label{Abel_Plana_2}
 \begin{aligned}
  \sum_{n=0}^{\infty} F(n+1/2) = \int_0^{\infty} F(t)dt - i\int_0^{\infty} \frac{dt}{e^{2\pi t}+1}
  \Big{[} F(it) - F(-it) \Big{]},
 \end{aligned}
\end{eqnarray}
with
\begin{eqnarray}
 \label{Mista_function_F_t}
 \begin{aligned}
  F(n+1/2) = \omega_{k,n}e^{-j\beta \omega_{k,n}} \ \ \ \text{where } \ \ \  \omega_{k,n} = 
  \sqrt{\frac{1}{(1+\lambda)}\Big{[} k^2 + m^2 + \Big{(}\Big{(}n+\frac{1}{2}\Big{)}\frac{\pi}{a}\Big{)}^2 \Big{]}},
 \end{aligned}
\end{eqnarray}
one expresses the energy as
\begin{eqnarray}
 \label{Mista_energy_temperatura_t_3}
 \begin{aligned}
 E_T = \frac{(1+\lambda)L^2}{2\pi} \sum_{j=1}^{\infty} e^{j\beta \mu} \int_0^{\infty} dkk \Bigg{\{} \int_0^{\infty} f(t)dt 
 -i\int_0^{\infty} \frac{dt}{e^{2\pi t}+1} \Big{[}F(it)-F(-it)\Big{]} \Bigg{\}}.
 \end{aligned}
\end{eqnarray}
The thermal Casimir energy is given by the second term on the right hand side of \eqref{Mista_energy_temperatura_t_3}. It reads,
\begin{eqnarray}
 \label{Mista_energy_Casimir_t_0}
 \begin{aligned}
 E_{T,C} &= - \frac{(1+\lambda)L^2}{2\pi} i \sum_{j=1}^{\infty} e^{j\beta \mu} \int_0^{\infty} dkk \int_0^{\infty} 
 \frac{dt}{e^{2\pi t}+1} \Big{[}F(it)-F(-it)\Big{]}.
 \end{aligned}
\end{eqnarray}
Considering the expressions \eqref{Mista_function_F_t} and performing the changes of variables $x^2=k^2+m^2$ and $u=\frac{t\pi}{a}$, we get
\begin{eqnarray}
 \label{Mista_energy_Casimir_t_1}
 E_{T,C} &=& - \frac{\sqrt{1+\lambda}}{2\pi^2} L^2 a i \sum_{j=1}^{\infty} e^{j\beta \mu} \int_m^{\infty} dxx \int_0^{\infty} 
 du \nonumber\\ &\times &
 \frac{\sqrt{x^2+(iu)^2}e^{-\frac{j\beta}{\sqrt{(1+\lambda)}} \sqrt{x^2+(iu)^2}} - \sqrt{x^2+(-iu)^2}e^{-\frac{j\beta}
 {\sqrt{(1+\lambda)}} \sqrt{x^2+(-iu)^2}}}{e^{2au}+1}  \   .
\end{eqnarray}

In the development of the integral over the variable $u$ we must consider two subintervals: the first one is $[0,x]$, and the second is $[x,\infty)$. It follows from \eqref{ident1} that the integral in the interval $[0,x]$ vanishes, so it remains to study only the 
integral in the second interval, consequently we get
\begin{eqnarray}
 \label{Mista_energy_Casimir_t_2}
 \begin{aligned}
 E_{T,C} = \frac{\sqrt{1+\lambda}}{\pi^2} L^2 a \sum_{j=1}^{\infty} e^{j\beta \mu} \int_m^{\infty} dxx \int_x^{\infty} 
 du \frac{\sqrt{u^2-x^2}}{e^{2au}+1}\cos \Bigg{(} \frac{j\beta}{\sqrt{1+\lambda}} \sqrt{u^2-x^2} \Bigg{)}.
 \end{aligned}
\end{eqnarray}

Changing the integral coordinate conveniently, we obtain
\begin{eqnarray}
 \label{Mista_energy_Casimir_t_3}
 \begin{aligned}
 E_{T,C} = \frac{\sqrt{1+\lambda}}{\pi^2} \frac{L^2}{a^3} \sum_{j=1}^{\infty} e^{j\beta \mu} \int_{am}^{\infty} dzz^3 
 \int_1^{\infty} dy \frac{\sqrt{y^2-1}}{e^{2zy}+1}\cos \Bigg{(} \frac{j\beta}{a\sqrt{1+\lambda}}z \sqrt{y^2-1} \Bigg{)}  \   .
 \end{aligned}
\end{eqnarray}
It is shown in Appendix \ref{appB}, that the integral above can be expressed in terms of the modified Bessel function of the second kind, $K_{\nu}(z)$, as
\begin{eqnarray}
 \label{Mista_Casimir_energy_t_4}
 \begin{aligned}
 E_{T,C} = - \frac{\sqrt{1+\lambda} L^2 m^2}{\pi^2 a} \sum_{j=1}^{\infty} e^{j\beta \mu} \frac{\partial}{\partial 
 \alpha_j} \Bigg{[} \alpha_j \sum_{n=1}^{\infty} \frac{(-1)^n}{\alpha_j^2+4n^2}K_2\Big{(}am\sqrt{\alpha_j^2+4n^2} \Big{)}
 \Bigg{]},
 \end{aligned}
\end{eqnarray}
where
\begin{eqnarray}
 \begin{aligned}
  \alpha_j = \frac{j \beta}{a\sqrt{1+\lambda}}.
 \end{aligned}
\end{eqnarray}

Here, also there is no closed expression for the sum over $n$ and $j$. The only way to provide some informations about the behavior of $E_{T,C}$ above, consists in 
consider the limit cases: $am \gg 1$, $am = 0$ and $am \ll 1$.
\begin{itemize}
 \item For case $am \gg 1$, the dominant term is $n=1$. Following the standard procedure, $E_{T,C}$ is given by 
 \begin{eqnarray}
 \label{Mista_energy_Casimir_t_5}
 \begin{aligned}
 \frac{E_{T,C}}{L^2} \approx \frac{\sqrt{1+\lambda} m^{3/2}}{2 \sqrt{2} \pi^{3/2} a^{3/2}} \sum_{j=1}^{\infty} \Bigg{[}
 \frac{8-\alpha_j^2\Big{(}3+2am\sqrt{\alpha_j^2+4}\Big{)}}{(\alpha_j^2+4)^{9/4}}\Bigg{]} e^{-am\sqrt{\alpha_j^2+4}+j\beta \mu} .
 \end{aligned}
\end{eqnarray}
Considering the low temperature limit, $\beta/a \gg 1$, we get 
\begin{eqnarray}
 \label{Mista_energy_Casimir_t_6}
 \begin{aligned}
 \frac{E_{T,C}}{L^2} \approx - \frac{(1+\lambda)^{5/4} a m^{5/2}}{\sqrt{2} \pi^{3/2} \beta^{3/2}} e^{-\beta \Big{(}
 \frac{m}{\sqrt{1+\lambda}}-\mu\Big{)}} .
 \end{aligned}
\end{eqnarray}
Notice that in this case, the Casimir energy per area unit decays exponentially with $\beta m$.
\end{itemize}

\begin{itemize}
 \item For case $am = 0$, we have to take  $\mu = 0$. In this case we get the closed expression below,
 \begin{eqnarray}
  \label{Mista_energy_Casimir_t_7}
  \begin{aligned}
   \frac{E_{T,C}}{L^2} = - \frac{\sqrt{1+\lambda}}{16 \pi^2 a^3} \sum_{j=1}^{\infty} \frac{1}{\alpha_j^4} & \Bigg{[} 48 + 
   \frac{1}{2} \pi \alpha_j \text{csch}^3 \Big{(}\frac{\pi \alpha_j}{2}\Big{)} \Big{[} 8 -3\pi^2 \alpha_j^2-(8+\pi^2\alpha_j^2)
   \\ & \times \text{cosh}(\pi \alpha_j) - 4\pi \alpha_j \text{sinh}(\pi \alpha_j) \Big{]} \Bigg{]} \  .
  \end{aligned}
 \end{eqnarray}
At low-temperature limit, and after performing the summation over $j$, we get,
 \begin{eqnarray}
  \label{Mista_energy_Casimir_t_8}
  \begin{aligned}
   \frac{E_{T,C}}{L^2} \approx - \frac{\pi^2 (1+\lambda)^{5/2} a}{30 \beta ^4} .
  \end{aligned}
 \end{eqnarray}

\end{itemize}

\begin{itemize}
 \item For case $am \ll 1$, the expression \eqref{Mista_Casimir_energy_t_4} is not convenient to analyze this limit. We have to take a more convenient expression to express. It is given by \eqref{sum1} considering $\alpha=\pi$ and adapting the other parameters to our case. Defining a new parameter $\sigma$ as 
 \begin{eqnarray}
 \label{sigma_parameter_mix_t}
 \begin{aligned}
 \alpha_j = j \sigma , \ \ \ \ \text{with} \ \ \ \ \sigma = \frac{\beta}{a\sqrt{1+\lambda}} \ ,
 \end{aligned}
 \end{eqnarray}
 and assuming the low-temperature limit, we can write\footnote{The details of this calculations are explicitly presented in Appendix \ref{appB}. }
 \begin{eqnarray}
 \label{Mista_energy_Casimir_t_9}
 \begin{aligned}
 \frac{E_{T,C}}{L^2} \approx - \frac{(1+\lambda) m^2}{2\pi^2 \beta} \sum_{j=1}^{\infty} e^{j \beta \mu} \Bigg{[} am 
 \frac{K_1(jam \sigma)}{j} + \frac{3}{\sigma} \frac{K_2(jam \sigma)}{j^2} \Bigg{]} \  .
 \end{aligned}
\end{eqnarray}
At this point, we will consider two possible cases: $m\beta > 1$ and $m\beta < 1$.
\begin{itemize}
 \item For case $m\beta > 1$, we get
 \begin{eqnarray}
  \label{Mista_energy_Casimir_t_10}
  \begin{aligned}
   \frac{E_{T,C}}{L^2} \approx - \frac{(1+\lambda)^{5/4} a}{2 \sqrt{2} \pi^{3/2}} \Bigg{[} \frac{m^{5/2}}{\beta^{3/2}} + 
   \frac{3\sqrt{1+\lambda}m^{3/2}}{\beta^{5/2}} \Bigg{]} e^{-\beta\Big{(}\frac{m}{\sqrt{1+\lambda}} -\mu\Big{)}} .
  \end{aligned}
 \end{eqnarray}
 \item For case $m\beta < 1$ ($\mu = 0$), we get
 \begin{eqnarray}
  \label{Mista_energy_Casimir_t_11}
  \begin{aligned}
   \frac{E_{T,C}}{L^2} \approx - \frac{ (1+\lambda )^{5/2} a }{120 \beta^4} \Bigg{[} 4\pi^2 + \frac{5 m^2 \beta^2}{(1+\lambda)} 
   \Bigg{]}.
  \end{aligned}
 \end{eqnarray}
 Where we can see a correction on $E_{T,C}$ of order $(m\beta)^2$ when compared with \eqref{Mista_energy_Casimir_t_8}.  
\end{itemize}
\end{itemize}

For this specific situation, the leading contribution to the total Casimir energy for the massless field, can be provided by using the corresponding result of \cite{Cruz:2017kfo}:
\begin{equation}
\frac{E_C}{L^2} \approx \frac{(1+\lambda)^{1/2}\pi^2}{30a^3} \Big{(} \frac{7}{384} - \frac{(1+\lambda)^2a^4}{\beta^4} \Big{)}.
\end{equation}
As we can see the Lorentz violating parameter appears modifying both contributions to the energy; moreover the thermal correction in this case is of order $O(Ta)^4$, so  weaker than in the previous boundary conditions obeyed by the field.

\subsubsection{Spacelike vector case}

For this case we have three different directions for the 4-vector $u^{\mu}$. They are $u^{(1)}=(0,1,0,0)$, $u^{(2)}=(0,0,1,0)$ and  $u^{(3)}=(0,0,0,1)$. The results obtained for Casimir energy  in two first cases are the same. So, let us 
concentrate on the vector $u^{(1)}=(0,1,0,0)$.

The Hamiltonian operator, $\hat{H}$, reads now
\begin{eqnarray}
 \label{Mista_Hamiltonian_operator_x}
 \begin{aligned}
 \hat{H} = \frac{1}{2} \int d^2{\vec{k}} \sum_{n=0}^{\infty} \omega_{\vec{k},n} \Bigg{[} 2\hat{a}_n^{\dagger}(\vec{k})
 \hat{a}_n(\vec{k}) + \frac{L^2}{(2\pi)^2} \Bigg{]} 
 \end{aligned}
\end{eqnarray}
and the dispersion relation is
\begin{eqnarray}
 \label{Mista_dispersion_x_0}
 \begin{aligned}
  \omega_{\vec{k},n} = \sqrt{(1-\lambda)k_x^2 + k_y^2 + \Big{[} \Big{(} n + \frac{1}{2} \Big{)}^2\frac{\pi}{a} \Big{]}^2 + m^2}.
 \end{aligned}
\end{eqnarray}
Consequently, the thermal correction to the Casimir energy is given by
\begin{eqnarray}
 \label{Mista_energy_temperatura_x_0}
 \begin{aligned}
  E_T = \int d^2 \vec{k} \sum_{n=0}^{\infty} \omega_{\vec{k},n} \langle \hat{a}_n^{\dagger}(\vec{k})\hat{a}_n(\vec{k}) \rangle  \  .
 \end{aligned}
\end{eqnarray}
Using the relations \eqref{occupation_numbers} and \eqref{relation_1}, we get
\begin{eqnarray}
 \label{Mista_energy_x_1} 
 \begin{aligned}
  E_T = \frac{L^2}{(2\pi)^2} \sum_{j=1}^{\infty} e^{j\beta \mu} \int d^2 \vec{k} \sum_{n=0}^{\infty} \omega_{\vec{k},n}
  e^{-j \beta \omega_{\vec{k},n}}.
 \end{aligned}
\end{eqnarray}
Developing the summation on $n$ by using \eqref{Abel_Plana_2}, and performing a change of coordinates $(k_x,k_y)$ to the polar
coordinates, with
\begin{eqnarray}
 \label{Mista_function_F_x}
 \begin{aligned}
  F \Big{(}n+1/2\Big{)} = \omega_{k,n}e^{-j \beta \omega_{k,n}}  \ \ \ \  \text{with}  \ \ \ \
  \omega_{k,n} = \sqrt{k^2 + m^2 + \Big{[}\Big{(}n+\frac{1}{2}\Big{)}\frac{\pi}{a}\Big{]}^2},
 \end{aligned}
\end{eqnarray}
we obtain
\begin{eqnarray}
 \label{Mista_energy_x_2}
  E_T &=& \frac{L^2}{2\pi \sqrt{1-\lambda}} \sum_{j=1}^{\infty} e^{j\beta \mu} \int_0^{\infty} dkk \Bigg{\{} \int_0^{\infty} 
  F(t)dt \nonumber\\
   &-& i \int_0^{\infty} \frac{dt}{e^{2\pi t}+1} \Big{[} F(it) - F(-it) \Big{]} \Bigg{\}}.
\end{eqnarray}

As we have already mentioned the thermal correction to the  Casimir energy is given by,
\begin{eqnarray}
 \label{Mista_energy_Casimit_x_0}
 \begin{aligned}
  E_{T,C} = - \frac{L^2}{2\pi \sqrt{1-\lambda}} i \sum_{j=1}^{\infty} e^{j\beta \mu} \int_0^{\infty} dkk \int_0^{\infty} 
  \frac{dt}{e^{2\pi t}+1} \Big{[} F(it) - F(-it) \Big{]}  \  .
 \end{aligned}
\end{eqnarray}
Considering the definition \eqref{Mista_function_F_x}, performing a change of variables $x^2=k^2+m^2$ and $u=\frac{\pi t}{a}$, we get
\begin{eqnarray}
 \label{Mista_energy_Casimit_x_1}
  E_{T,C} &=& - \frac{aL^2}{2\pi^2 \sqrt{1-\lambda}} i \sum_{j=1}^{\infty} e^{j\beta \mu} \int_m^{\infty} dxx \int_0^{\infty} du\nonumber\\
  &\times& \frac{\sqrt{x^2+(iu)^2}e^{-j\beta \sqrt{x^2+(iu)^2}} - \sqrt{x^2+(-iu)^2}e^{-j\beta \sqrt{x^2+(-iu)^2}}} 
  {e^{2au}+1}  \  .
\end{eqnarray}

Again we consider the integral over the variable $u$ in two subintervals: $[0, x]$ and the $[x, \infty)$. It follows from \eqref{ident1} that the integral in the segment $[0, x]$ vanishes, so it remains  to take the integral in the interval $[x, \infty)$. Performing the change of variables, $u=xy$ and $z=ax$, we arrive at
\begin{eqnarray}
 \label{Mista_energy_Casimir_x_2}
 \begin{aligned}
  E_{T,C} &= \frac{L^2}{\pi^2 \sqrt{1-\lambda} a^3} \sum_{j=1}^{\infty} e^{j\beta \mu} \int_{am}^{\infty} dzz^3 
  \int_1^{\infty} dy \frac{\sqrt{y^2-1}}{e^{2zy}+1} \cos \Bigg{(} \frac{j\beta}{a} z\sqrt{y^2-1} \Bigg{)}  \  .
 \end{aligned}
\end{eqnarray}
Expressing the above integral in terms of the function $K_\nu(z)$, we obtain,
\begin{eqnarray}
 \label{Mista_Casimir_energy_x_3}
 \begin{aligned}
 E_{T,C} = - \frac{L^2  m^2}{\pi^2 \sqrt{1-\lambda} a} \sum_{j=1}^{\infty} e^{j\beta \mu} \frac{\partial}{\partial 
 \alpha_j} \Bigg{[} \alpha_j \sum_{n=1}^{\infty} \frac{(-1)^n}{\alpha_j^2+4n^2}K_2\Big{(}am\sqrt{\alpha_j^2+4n^2} \Big{)}
 \Bigg{]},
 \end{aligned}
\end{eqnarray}
where
\begin{eqnarray}
 \begin{aligned}
  \alpha_j = \frac{j \beta}{a}.
 \end{aligned}
\end{eqnarray}

As we have already mentioned, there is no closed expression for the summation over $n$. The only way to provide some informations about the behavior  of \eqref{Mista_Casimir_energy_x_3} consists in considering the limits cases: $am \gg 1$, $am = 0$ and $am \ll 1$.
\begin{itemize}
 \item For case $am \gg 1$, the dominant term is for $n=1$ which gives 
 \begin{eqnarray}
 \label{Mista_energy_Casimir_x_4}
 \frac{E_{T,C}}{L^2} &\approx& \frac{m^{3/2}}{2 \sqrt{2} \pi^{3/2} \sqrt{1-\lambda}a^{3/2}} \sum_{j=1}^{\infty} \Bigg{[}
 \frac{8-\alpha_j^2\Big{(}3+2am\sqrt{\alpha_j^2+4}\Big{)}}{(\alpha_j^2+4)^{9/4}}\Bigg{]}
 \nonumber\\
 &\times& e^{-am\sqrt{\alpha_j^2+4}+j\beta \mu}  \ .
\end{eqnarray}
Considering the low temperature limit, $\beta/a \gg 1$, we get
\begin{eqnarray}
 \label{Mista_energy_Casimir_x_5}
 \begin{aligned}
 \frac{E_{T,C}}{L^2} \approx - \frac{a m^{5/2}}{\sqrt{2} \pi^{3/2} \sqrt{1-\lambda} \beta^{3/2}} e^{-\beta (m-\mu)} .
 \end{aligned}
\end{eqnarray}
Notice that in this case, the thermal Casimir energy per area unit decays exponentially with $\beta m$.
 \item For case $am = 0$ ($\mu = 0$), we get the closed expression
 \begin{eqnarray}
  \label{Mista_energy_Casimir_x_6}
   \frac{E_{T,C}}{L^2} &=& - \frac{1}{16 \pi^2 \sqrt{1-\lambda} a^3} \sum_{j=1}^{\infty} \frac{1}{\alpha_j^4}  \Bigg{[} 48 + 
   \frac{1}{2} \pi \alpha_j \text{csch}^3 \Big{(}\frac{\pi \alpha_j}{2}\Big{)} \Big{[} 8 -3\pi^2 \alpha_j^2\nonumber\\
    &-&(8+\pi^2\alpha_j^2) \text{cosh}(\pi \alpha_j) - 4\pi \alpha_j \text{sinh}(\pi \alpha_j) \Big{]} \Bigg{]} \   .
 \end{eqnarray}
Considering now the low-temperature limit, we obtain
 \begin{eqnarray}
  \label{Mista_energy_Casimir_x_7}
  \begin{aligned}
   \frac{E_{T,C}}{L^2} \approx -\frac{\pi^2 a}{30 \sqrt{1-\lambda} \beta^4 } .
  \end{aligned}
 \end{eqnarray}
 \item For the case $am \ll 1$, and after some intermediate steps already mentioned, we get
 \begin{eqnarray}
 \label{Mista_energy_Casimir_x_8}
 \begin{aligned}
 \frac{E_{T,C}}{L^2} \approx - \frac{m^2}{2 \pi^2 \sqrt{1-\lambda } \beta } \sum_{j=1}^{\infty} e^{j\beta \mu} \Bigg{[}
 am \frac{K_1(jam\sigma)}{j} + \frac{3}{\sigma} \frac{K_2(jam\sigma)}{j^2} \Bigg{]}  \  .
 \end{aligned}
\end{eqnarray}
Where we have used the definition  
\begin{eqnarray}
\label{sigma_parameter_mix_x}
\begin{aligned}
\alpha_j = j \sigma , \ \ \ \ \text{with} \ \ \ \ \sigma = \frac{\beta}{a} .
\end{aligned}
\end{eqnarray}
At this point, we will consider two  possible situations: $m\beta > 1$ and $m\beta < 1$.

\begin{itemize}
 \item For case $m\beta > 1$, we get
 \begin{eqnarray}
  \label{Mista_energy_Casimir_x_9}
  \begin{aligned}
   \frac{E_{T,C}}{L^2} \approx - \frac{a}{2\sqrt{2}\pi^{3/2}\sqrt{1-\lambda}} \Bigg{[} \frac{3 m^{3/2}}{\beta^{5/2}} + 
   \frac{m^{5/2}}{\beta^{3/2}} \Bigg{]} e^{-\beta(m-\mu)} .
  \end{aligned}
 \end{eqnarray}
 \item For case $m\beta < 1$ ($\mu = 0$), we get
 \begin{eqnarray}
  \label{Mista_energy_Casimir_x_10}
  \begin{aligned}
   \frac{E_{T,C}}{L^2} \approx - \frac{a}{120 \sqrt{1-\lambda} \beta^4} \Bigg{[} 4\pi^2 - 5 m^2 \beta^2 \Bigg{]} .
  \end{aligned}
 \end{eqnarray}
 We can see that there appears in the expression above, a corrections proportional to $(m\beta)^2$, when compared with the massless case, Eq. \eqref{Mista_energy_Casimir_x_7}
\end{itemize}
\end{itemize}

Below we present the leading contribution to the total Casimir energy for the massless field:
\begin{equation}
\frac{E_C}{L^2} \approx \frac{(1-\lambda)^{-1/2}\pi^2}{30a^3} \Big{(}\frac{7}{384}-\frac{a^4}{\beta^4} \Big{)}. 
\end{equation}
Here also the thermal correction is of order $O(Ta)^4$, and the modification on the total Csimir energy due to the parameter $\lambda$ is just a multiplicative factor.

Finally, we now consider the 4-vector orthogonal to the plates:
\begin{eqnarray}
 \begin{aligned}
  u^{(3)}=(0,0,0,1).
 \end{aligned}
\end{eqnarray}
The Hamiltonian operator, $\hat{H}$, remains the same
\begin{eqnarray}
 \label{Mista_Hamiltonian_operator_z}
 \begin{aligned}
  \hat{H} = \frac{1}{2} \int d^2\vec{k} \sum_{n=0}^{\infty} \omega_{\vec{k},n} \Bigg{[} 2\hat{a}_n^{\dagger}(\vec{k})
  \hat{a}_n(\vec{k}) + \frac{L^2}{(2\pi)^2} \Bigg{]} \   .
 \end{aligned}
\end{eqnarray}
However, for this case the dispersion relation is modified to
\begin{eqnarray}
 \label{Mista_dispersion_relation_z}
 \begin{aligned}
  \omega_{\vec{k},n} = \sqrt{k_x^2 + k_y^2 + (1-\lambda) \Big{[}\Big{(} n+\frac{1}{2} \Big{)}\frac{\pi}{a}\Big{]}^2 +m^2}.
 \end{aligned}
\end{eqnarray}

Consequently, the thermal correction for energy is
\begin{eqnarray}
 \label{Mista_energy_temperatura_z_0}
 \begin{aligned}
  E_T = \int d^2\vec{k} \sum_{n=0}^{\infty} \omega_{\vec{k},n} \langle \hat{a}_n^{\dagger}(\vec{k})\hat{a}_n(\vec{k}) \rangle .
 \end{aligned}
\end{eqnarray}
Using the thermal occupation number \eqref{occupation_numbers} and the relation \eqref{relation_1} we find, 
\begin{eqnarray}
 \label{Mista_energy_temperatura_z_2}
 \begin{aligned}
  E_T = \frac{L^2}{(2\pi)^2} \sum_{j=1}^{\infty} e^{j \beta \mu} \int d^2\vec{k} \sum_{n=0}^{\infty} \omega_{\vec{k},n}
  e^{-j\beta \omega_{\vec{k},n}} .
 \end{aligned}
\end{eqnarray}
Performing a change of variables in the Cartesian coordinates $(k_x,k_y)$ to polar one $(k,\theta)$, with $k=\sqrt{k^2_x+k_y^2}$, the integral on angular variable is promptly done. Finally using the Abel-Plana formula \eqref{Abel_Plana_2}, we  get
\begin{eqnarray}
 \label{Mista_energy_temperatura_z_3}
 \begin{aligned}
  E_T = \frac{L^2}{2\pi} \sum_{j=1}^{\infty} e^{j \beta \mu} \int_0^{\infty} dkk \Bigg{\{} \int_0^{\infty} F(t)df 
  -i \int_0^{\infty} \frac{dt}{e^{2\pi t}+1} \Big{[} F(it) - F(-it) \Big{]} \Bigg{\}},
 \end{aligned}
\end{eqnarray}
where
\begin{eqnarray}
 \label{Mista_function_F_z}
 \begin{aligned}
  F \Big{(} n+1/2 \Big{)} = \omega_{k,n}e^{-j \beta \omega_{k,n}} \ \ \ \ \text{ with} \ \ \ \  
  \omega_{k,n} = \sqrt{k^2 + m^2 + (1-\lambda)\Big{(}\Big{(}n+\frac{1}{2}\Big{)}\frac{\pi}{a}\Big{)}^2}.
 \end{aligned}
\end{eqnarray}

Therefore the thermal correction for Casimir energy is given by
\begin{eqnarray}
 \label{Mista_energy_Casimir_z_0}
 \begin{aligned}
  E_{T,C} = - \frac{L^2}{2\pi} i \sum_{j=1}^{\infty} e^{j \beta \mu} \int_0^{\infty} dkk \int_0^{\infty} \frac{dt}{e^{2\pi t}+1} 
  \Big{[} F(it) - F(-it) \Big{]} \  .
 \end{aligned}
\end{eqnarray}
Defining  new variables $x^2=k^2+m^2$ and $u=\frac{\pi t}{b}$ where 
$\frac{1}{b}=\frac{\sqrt{1-\lambda}}{a}$, the following expression is obtained for the  energy:
\begin{eqnarray}
 \label{Mista_energy_Casimir_z_1}
  E_T& =& - \frac{bL^2}{2\pi^2} i \sum_{j=1}^{\infty} e^{j \beta \mu} \int_m^{\infty} dxx \int_0^{\infty} du \nonumber\\ 
  &\times&  \frac{\sqrt{x^2+(iu)^2}e^{-j \beta \sqrt{x^2+(iu)^2}} - \sqrt{x^2+(-iu)^2}e^{-j \beta \sqrt{x^2+(-iu)^2}}}
  {e^{2bu}+1}.
\end{eqnarray}

Once more we have to divide the integral over the variable $u$ in two subintervals: $[0, x]$ and the $[x, \infty)$. It follows from \eqref{ident1} that the integral in the interval $[0, x]$ vanishes, so it remains the integral in the segment $[x, \infty)$. Performing following change of variables, $u = xy$ and $z = bx$, we arrive at 
\begin{eqnarray}
 \label{Mista_energy_Casimir_z_2}
 \begin{aligned}
  E_{T,C} = \frac{L^2}{\pi^2b^3} \sum_{j=1}^{\infty} e^{j \beta \mu} \int_{bm}^{\infty} dzz^3 \int_1^{\infty} dy
  \frac{\sqrt{y^2-1}}{e^{2zy}+1} \cos \Bigg{(} \frac{j \beta}{b} z \sqrt{y^2-1} \Bigg{)},
 \end{aligned}
\end{eqnarray}
As before the above integral can be writing in term of modified Bessel function of second kind, $K_{\nu}(z)$ as shown below,
\begin{eqnarray}
 \label{Mista_energy_Casimir_z_3}
 \begin{aligned}
  E_{T,C} = - \frac{L^2 m^2}{\pi^2 b} \sum_{j=1}^{\infty} e^{j \beta \mu} \frac{\partial}{\partial \alpha_j} \Bigg{[} \alpha_j
  \sum_{n=1}^{\infty} \frac{(-1)^n}{\alpha_j^2+4n^2} K_2 \Big{(} bm \sqrt{\alpha_j^2+4n^2} \Big{)} \Bigg{]},
 \end{aligned}
\end{eqnarray}
where
\begin{eqnarray}
 \begin{aligned}
  \alpha_j = \frac{j \beta}{b}.
 \end{aligned}
\end{eqnarray}

Taking the asymptotic limits, $am \gg 1$, $am = 0$ and $am \ll 1$ we get:
\begin{itemize}
 \item For case $am \gg 1$, the dominant term is $n=1$ given by 
 \begin{eqnarray}
 \label{Mista_energy_Casimir_z_4}
 \begin{aligned}
 \frac{E_{T,C}}{L^2} \approx  \frac{m^{3/2}}{2\sqrt{2}\pi^{3/2} b^{3/2}} \sum_{j=1}^{\infty} \Bigg{[}
 \frac{8-\alpha_j^2\Big{(}3+2bm\sqrt{\alpha_j^2+4}\Big{)}}{(\alpha_j^2+4)^{9/4}}\Bigg{]} e^{-bm\sqrt{\alpha_j^2+4}+j\beta \mu} .
 \end{aligned}
\end{eqnarray}
Considering the low temperature limit $\beta/a \gg 1$, it is possible to develop the summation over $j$, and we obtain
\begin{eqnarray}
 \label{Mista_energy_Casimir_z_5}
 \begin{aligned}
 \frac{E_{T,C}}{L^2} \approx - \frac{a m^{5/2}}{\sqrt{2} \pi^{3/2} \sqrt{1-\lambda} \beta^{3/2}} e^{-\beta (m-\mu)} .
 \end{aligned}
\end{eqnarray}
Notice that in this case, the Casimir energy per area unit decays exponentially with $\beta m$.
 \item For case $am = 0$ ($\mu = 0$), we obtain a closed expression
 \begin{eqnarray}
  \label{Mista_energy_Casimir_z_6}
   \frac{E_{T,C}}{L^2} &=& - \frac{1}{16 \pi^2 b^3} \sum_{j=1}^{\infty} \frac{1}{\alpha_j^4} \Bigg{[} 48 + 
   \frac{1}{2} \pi \alpha_j \text{csch}^3 \Big{(}\frac{\pi \alpha_j}{2}\Big{)} \Big{[} 8 -3\pi^2 \alpha_j^2\nonumber\\
   &-&(8+\pi^2\alpha_j^2)\text{cosh}(\pi \alpha_j) - 4\pi \alpha_j \text{sinh}(\pi \alpha_j) \Big{]} \Bigg{]} .
 \end{eqnarray}
Considering now the low-temperature limit, we get
 \begin{eqnarray}
  \label{Mista_energy_Casimir_z_7}
  \begin{aligned}
   \frac{E_{T,C}}{L^2} \approx - \frac{\pi^2 a}{30 \sqrt{1-\lambda} \beta^4} .
  \end{aligned}
 \end{eqnarray}
 \item For case $am \ll 1$, we adopt the procedure already exhibited, and we arrive, at low-temperature limit, to
 \begin{eqnarray}
 \label{Mista_energy_Casimir_z_8}
 \begin{aligned}
 \frac{E_{T,C}}{L^2} \approx - \frac{m^2}{2 \pi^2 \beta } \sum_{j=1}^{\infty} e^{j\beta \mu} \Bigg{[}
 bm \frac{K_1(jbm\sigma)}{j} + \frac{3}{\sigma} \frac{K_2(jbm\sigma)}{j^2} \Bigg{]} \ .
 \end{aligned}
\end{eqnarray}
Being $\sigma$ defined as
\begin{eqnarray}
\label{sigma_parameter_mix_z}
\begin{aligned}
\alpha_j = j \sigma , \ \ \ \ \text{with} \ \ \ \ \sigma = \frac{\beta}{b} .
\end{aligned}
\end{eqnarray}
At this point, we will consider two possible cases: $m\beta > 1$ and $m\beta < 1$.

\begin{itemize}
\item For case $m\beta > 1$, we get
 \begin{eqnarray}
  \label{Mista_energy_Casimir_z_9}
  \begin{aligned}
   \frac{E_{T,C}}{L^2} \approx - \frac{a}{2\sqrt{2}\pi^{3/2}\sqrt{1-\lambda}} \Bigg{[} \frac{3m^{3/2}}{\beta^{5/2}} + 
   \frac{m^{5/2}}{\beta^{3/2}} \Bigg{]} e^{-\beta(m-\mu)} .
  \end{aligned}
 \end{eqnarray}
 \item For case $m\beta < 1$ ($\mu = 0$), we get
 \begin{eqnarray}
  \label{Mista_energy_Casimir_x_10a}
  \begin{aligned}
   \frac{E_{T,C}}{L^2} \approx - \frac{a}{120 \sqrt{1-\lambda} \beta^{4}} \Big{[}4\pi^2-5m^2 \beta^2 \Big{]} .
  \end{aligned}
 \end{eqnarray}

\end{itemize}
\end{itemize}
To complete this subsection, below we present the leading contribution to the total Casimir energy for the massless field:
\begin{equation}
\frac{E_C}{L^2} \approx  \frac{(1-\lambda)^{3/2}\pi^2}{30a^3} \Big{(}\frac{7}{384}-\frac{(1-\lambda)^2a^4}{\beta^4} \Big{)} .
\end{equation}

\section{Concluding Remarks}

In this paper we have considered the Casimir effect in the Lorentz-breaking CPT-even extension of the scalar field theory at the finite temperature, considering a non-vanishing chemical potential $\mu$ for the bosonic field. We treated situations of different possible directions of the Lorentz-breaking constant vector $u^{\mu}$ for  different boundary conditions obeyed by the field on the parallel plates, that is, Dirichlet, von Neumann and mixed ones. We found that the Casimir energy can be expressed in terms of modified Bessel function, $K_\nu(z)$, as presented by  \eqref{Dirichlet_Casimir_energy_t_5}, \eqref{Dirichlet_Casimir_energy_x_4} and \eqref{Dirichlet_Casimir_energy_temperature_z_4}, for Dirichlet and Newman conditions, and by \eqref{Mista_Casimir_energy_t_4} for mixed one. Since there is no closed expressions for the corresponding Casimir energy, we only can provide some informations considering specific limits of mass and temperature. 

By the above mentioned expressions, the thermal Casimir energy crucially depends on the dimensionless parameter $M=am$, being $a$ the distance between plates, and $m$ the mass of the field, and on $\beta/a=1/{Ta}$, which we consider much bigger than unity. For $am>>1$, we could show that energy decays exponentially, which is rather similar to the zero temperature situation. For massless case, and taking $\mu=0$, there appear third and fourth order corrections in temperature for the Dirichlet and Neumann boundary condition cases, and  a correction of fourth order in temperature for mixed boundary condition case. For $am<<1$, the expressions for the energy are rather cumbersome which does not allow to study this limit. So, we used the identity \eqref{sum1} adapting conveniently the parameters of the system under investigation. We founded that the thermal Casimir energy depends on the parameter $m\beta$. For the case $m\beta<<1$, we have to take $\mu=0$, so there appears a second order correction on $(m\beta)$ . 

In general, we have noticed that the dependence of the results on the Lorentz-breaking  parameter $\lambda$ occurs through the factor $(1\pm\lambda)^n$ with different signs and different values of $n$ for different 
boundary conditions multiplying either the whole result or some contributions to it, both zero-temperature and finite-temperature ones, thus, the Lorentz symmetry breaking modifies the impact of the finite temperature only through in a small way. Also, we note that in the low temperature limit, both Lorentz-breaking and finite-temperature impacts are small. 

A possible continuation of this study could consist in its realization in the case of other fields, in particular, the spinor field. We plan to perform this calculation in a forthcoming paper.


{\bf Acknowledgements.} We would like to thank A. A. Saharian for a pertinent discussion. This work was partially supported by Conselho
Nacional de Desenvolvimento Cient\'{\i}fico e Tecnol\'{o}gico (CNPq). A. Yu. P. has been partially supported by the CNPq 
through the project No. 303783/2015-0, E. R. Bezerra de Mello through the project 
No. 313137/2014-5. M. B. Cruz has been supported by  
Coordenação de Aperfeiçoamento de Pessoal de Nível Superior (CAPES).

\appendix
\section{Calculation of integral I}
\label{appA}

In this Appendix we present the procedure adopted to express the integral representation for the thermal Casimir energy, given by  \eqref{Dirichlet_Casimir_energy_t_4}, \eqref{Dirichlet_Casimir_energy_x_3} and 
\eqref{Dirichlet_Casimir_energy_temperature_z_3}, 
\begin{eqnarray}
 \label{appendix_Dirichlet_0}
 \begin{aligned}
  {\cal I}_j = \int_{M}^{\infty} dzz^3 \int_1^{\infty} dy \frac{\sqrt{y^2 - 1}}{e^{2zy}-1} \cos (\alpha_j z \sqrt{y^2 - 1}) \ ,
 \end{aligned}
\end{eqnarray}
in terms of a summation of modified Bessel function of the second kind, $K_\nu(z)$. In Eq. \eqref{appendix_Dirichlet_0} it is assumed that $M=ma$ and $\alpha_j=j\sigma=j\frac{\beta}{a}(1-\lambda)^\kappa$ being $\kappa=-1/2, \ 0$ and $1/2$.

We can write \eqref{appendix_Dirichlet_0} as
\begin{eqnarray}
 \label{appendix_Dirichlet_1}
 \begin{aligned}
  {\cal I}_j & = \frac{\partial}{\partial \alpha_j} \int_{M}^{\infty} dzz^2 \int_1^{\infty} dy \frac{\sin (\alpha_j z 
  \sqrt{y^2 - 1})}{e^{2zy}-1}  \\
  & = \frac{\partial}{\partial \alpha_j} \sum_{n=1}^{\infty} \int_M^{\infty} dzz^2 \int_1^{\infty} dy
  e^{-2nzy} \sin (\alpha_j z \sqrt{y^2 - 1}) .
 \end{aligned}
\end{eqnarray}
Performing the change of variable $x^2 = y^2 -1$ and  using the integral representation to the modified Bessel functions, $K_{\nu}(z)$, given in \cite{prudnikov:1986}, we get
\begin{eqnarray}
 \label{appendix_Dirichlet_integral_2}
 \begin{aligned}
  {\cal I}_j & = \frac{\partial}{\partial \alpha_j} \sum_{n=1}^{\infty} \int_M^{\infty} dzz^2 \int_0^{\infty} dx
  \frac{x \sin (\alpha_j z x) e^{-2nz\sqrt{x^2 + 1}}}{\sqrt{x^2 + 1}} \\
  & = \frac{\partial}{\partial \alpha_j} \sum_{n=1}^{\infty} \frac{\alpha_j}{\sqrt{\alpha_j^2+4n^2}} \int_M^{\infty} dzz^2 
  K_1\Big{(} z\sqrt{\alpha_j^2+4n^2} \Big{)}  \  .
 \end{aligned}
\end{eqnarray}
Moreover, doing the change of variable $x = z\sqrt{\alpha^2+4n^2}$, we get
\begin{eqnarray}
 \label{appendix_Dirichlet_integral_3}
 \begin{aligned}
  {\cal I}_j = \frac{\partial}{\partial \alpha_j} \sum_{n=1}^{\infty} \frac{\alpha_j}{(\alpha_j^2+4n^2)^2} 
  \int_{M\sqrt{\alpha_j^2+4n^2}}^{\infty} dxx^2 K_1(x) .
 \end{aligned}
\end{eqnarray}
Finally using the relation
\begin{eqnarray}
 \label{recorrence_relation}
 \begin{aligned}
  \frac{\partial}{\partial x} [x^{\nu} K_{\nu}(x)] = -x^{\nu}K_{\nu-1}(x),
 \end{aligned}
\end{eqnarray}
we get
\begin{eqnarray}
 \label{appendix_Dirichlet_integral_4}
 \begin{aligned}
  {\cal I}_j  = M^2 \frac{\partial}{\partial \alpha_j} \Bigg{[} \alpha_j \sum_{n=1}^{\infty} \frac{K_2\Big{(}M\sqrt{\alpha_j^2
  + 4n^2}\Big{)}}{\alpha_j^2+4n^2} 
  \Bigg{]}.
 \end{aligned}
\end{eqnarray}

Because there is no closed expression for the sum above, only two distinct limits can be considered: $M \gg 1$ and $M = 0$.
\begin{itemize}
 \item For the case $M \gg 1$, the dominant contribution comes from the term $n = 1$:
 \begin{eqnarray}
  \label{appendix_Dirichlet_integral_big_M_0}
  \begin{aligned}
   {\cal I}_j & \approx M^2 \frac{\partial}{\partial \alpha_j} \Bigg{[} \alpha_j \frac{K_2\Big{(}M\sqrt{\alpha_j^2+4}\Big{)}}
   {\alpha_j^2+4} \Bigg{]}   \  .
  \end{aligned}
 \end{eqnarray}
Also we use the asymptotic expansion for large arguments of the modified Bessel function \cite{abramowitz:1966}, $K_{\nu}(z)$, 
 \begin{eqnarray}
 \label{large_Bessel}
  K_{\nu}(z) \approx \sqrt{\frac{\pi}{2z}} e^{-z}  \  .
  \end{eqnarray}
  So we get:
  \begin{eqnarray}
  \label{Dirichlet_integral_M_g}
  {\cal I_j}& \approx &\sqrt{\frac{\pi}{2}} M^{\frac{3}{2}}  \frac{\partial}{\partial \alpha_j} \Bigg{[}  \frac{\alpha_j}{(\alpha_j^2+4)^{\frac{5}
  		{4}}} e^{-M\sqrt{\alpha_j^2+4}} \Bigg{]} \nonumber\\
  &\approx& \sqrt{\frac{\pi}{8}} M^{\frac{3}{2}} \frac{[8-\alpha_j^2(3+2M\sqrt{\alpha_j^2+4})]}{(\alpha_j^2+4)^{\frac{9}{4}}} 
  e^{-M\sqrt{\alpha_j^2+4}}  \   .
  \end{eqnarray}
  In low-temperature limit, $\alpha_j>>1$, we have
  \begin{eqnarray}
  \label{Integral_2_M_g_30}
  \begin{aligned}
  {\cal I}_j & \approx - \sqrt{\frac{\pi M^5}{2}} \frac{e^{-M \alpha_j}}{\alpha_j^{3/2}} \  .
  \end{aligned}
  \end{eqnarray}
  So we can develop the summation over $j$ in the final calculation for the thermal correction to the Casimir energy, $E_{T,C}.$
 \item For case $M = 0$, we using the asymptotic expansion for modified Bessel function \cite{abramowitz:1966}, $K_{\nu}(z)$, 
 for small arguments:
\begin{eqnarray}
 \label{small_Besseal}
 K_{\nu}(z) \approx \frac{1}{2} \Gamma(\nu) \Big{(} \frac{1}{2} z \Big{)}^{-\nu}  \  .
\end{eqnarray}
So we obtain,
\begin{eqnarray} \nonumber
\label{Dirichlet_integral_M_p_1}
{\cal I_j} \approx\frac18\frac{\partial}{\partial \alpha_j} \Bigg{[} \alpha_j \sum_{n=1}^{\infty} \frac{1}{(n^2+\alpha_j^2/4)^2}\Bigg{]}=-\frac14\frac{\partial^2}{\partial\alpha_j^2} \sum_{n=1}^{\infty} \frac{1}{n^2+\alpha_j^2/4}\  .
\end{eqnarray}
The summation on $n$ provides,
\begin{eqnarray}
\sum_{n=1}^\infty\frac1{n^2+\alpha_j^2/4}={\frac {\pi\alpha_j \,\coth \left( \frac{\pi\alpha_j}2\right) -2}{{\alpha_j}^{2}}} \  .
\end{eqnarray}
The next step is to derivate with respect to $\alpha_j$.  The result is
\begin{eqnarray}
\label{appendix_Dirichlet_integral_M_zero_1}
\begin{aligned}
{\cal I}_j = \frac{1}{8\alpha_j^4} \Bigg{[} 24 - 4 \pi \alpha_j \text{coth}\Big{(}\frac{\pi \alpha_j}{2}\Big{)} - 
\pi^2 \alpha^2_j \Big{[} 2 + \pi \alpha_j \text{coth}\Big{(}\frac{\pi \alpha_j}{2}\Big{)} \Big{]} \text{csch}^2\Big{(}
\frac{\pi \alpha_j}{2}\Big{)} \Bigg{]}.
\end{aligned}
\end{eqnarray}
Considering the parameter $\alpha_j$ large, we get\footnote{For $\alpha_j$ large, we have approximation of the 
	$\text{coth}\Big{(}\frac{\pi\alpha_j}{2}\Big{)}$ and $\text{csch}\Big{(}\frac{\pi\alpha_j}{2}\Big{)}$ by $1$ and $0$, 
	respectively.}
\begin{eqnarray}
\label{appendix_Dirichlet_integral_M_zero_0}
\begin{aligned}
{\cal I}_j \approx \frac{3}{\alpha_j^4} - \frac{\pi}{2\alpha_j^3}.
\end{aligned}
\end{eqnarray}
\end{itemize}

 For the case $M \ll 1$ the expression \eqref{appendix_Dirichlet_integral_4} is not convenient to analyze this limit. So, we will adopt another representation. Let us define a new function $f_{\nu}(z)$ by
 \begin{eqnarray}
 \label{function_f}
  \begin{aligned}
   f_{\nu}(z) = \frac{K_{\nu}(z)}{z^{\nu}} \  .
  \end{aligned}
 \end{eqnarray}
 In terms of this function, Eq. \eqref{appendix_Dirichlet_integral_4} reads,
\begin{eqnarray}
 \label{appendix_Dirichlet_small_M_0}
 \begin{aligned}
  {\cal I}_j & = M^4 \frac{\partial}{\partial \alpha_j}\Bigg{[} \alpha_j \sum_{n=1}^{\infty}f_2\Big{(}M\sqrt{\alpha_j^2+4n^2}
  \Big{)}\Bigg{]} \\
  & = M^4 \frac{\partial}{\partial \alpha_j} \Bigg{[} \frac{\alpha_j}{2} \sum_{n=-\infty}^{\infty} f_2\Big{(}M\sqrt{\alpha_j^2+
  4n^2}\Big{)} - \frac{\alpha_j}{2} f_2(M\alpha_j) \Bigg{]}  \  .
 \end{aligned}
\end{eqnarray}
From this point on, we proceed the summation on the right hand side of \eqref{appendix_Dirichlet_small_M_0} by using the identity  below \cite{deMello:2012xm}:
\begin{eqnarray}
\label{sum_relation}
\begin{aligned}
\sum_{n=-\infty}^{\infty} \cos(n \alpha) f_{\nu} (c \sqrt{b^2+a^2 n^2}) = \frac{\sqrt{2 \pi}}{a c^{2\nu}} \sum_{n=-\infty}^{\infty}
w_n^{2\nu-1} f_{\nu-1/2}(bw_n)
\end{aligned}
\end{eqnarray}
with $a,b,c > 0$ and $w_n = \sqrt{(2\pi n + \alpha)^2/a^2 + c^2}$. Considering $\alpha=0$, and adapting the other constants according to our problem, we obtain:
\begin{eqnarray}
 \label{appendix_Dirichlet_sum_0}
  \Sigma_{j} & =& \sum_{n=-\infty}^{\infty} f_2\Big{(}M\sqrt{\alpha_j^2+4n^2}\Big{)} \nonumber\\
  &=&\frac{\sqrt{2\pi}}{2 M^4} 
  \sum_{n=-\infty}^{\infty} (\pi^2 n^2 + M^2)^{3/2} f_{3/2}\Big{(}\alpha_j \sqrt{\pi^2 n^2 + M^2}\Big{)} \nonumber\\
  & = &\frac{\sqrt{2\pi}}{M^4 \alpha_j^{3/2}} \sum_{n=1}^{\infty} (\pi^2 n^2 + M^2)^{3/4} K_{3/2}\Big{(}\alpha_j \sqrt{\pi^2 
  n^2 + M^2}\Big{)} + \frac{\sqrt{2 \pi}}{2M^{5/2} \alpha_j^{3/2}} K_{3/2}(M \alpha_j) \ .\nonumber\\
\end{eqnarray}

Substituting \eqref{appendix_Dirichlet_sum_0} into \eqref{appendix_Dirichlet_small_M_0} we have
\begin{eqnarray}
 \label{appendix_Dirichlet_small_M_1}
  {\cal I}_j& =& \frac{\partial}{\partial \alpha_j} \Bigg{[} \frac{\sqrt{2\pi}}{2 \sqrt{\alpha_j}} \sum_{n=1}^{\infty}
  (\pi^2 n^2 + M^2)^{3/4} K_{3/2}(\alpha_j \sqrt{\pi^2 n^2 + M^2}) \nonumber\\
  &+& \frac{\sqrt{2\pi}M^{3/2}}{4 \sqrt{\alpha_j}}
  K_{3/2}(M \alpha_j) - \frac{M^2}{2 \alpha_j} K_{2}(M \alpha_j) \Bigg{]},
\end{eqnarray}
Remembering that the modified Bessel function \cite{abramowitz:1966}, $K_{3/2}(z)$, is given by
\begin{eqnarray}
 K_{3/2}(z) = \sqrt{\frac{\pi}{2}} \Big{(} 1 + \frac{1}{z} \Big{)} \frac{e^{-z}}{\sqrt{z}} \   ,
\end{eqnarray}
as $M \ll 1$, we can expand the expression below in powers of $M$, and procedure the summation in $n$. The dominate term is given below,
\begin{eqnarray}
 && \sum_{n=1}^{\infty} (\pi^2 n^2 + M^2)^{3/4} K_{3/2}(\alpha_j \sqrt{\pi^2 n^2 + M^2}) \nonumber\\
  && \approx - \sqrt{\frac{\pi}{8}} \Bigg{[} \frac{2 + M^2\alpha_j^2 e^{\pi \alpha_j} -2\pi \alpha_j e^{\pi \alpha_j} - 
  2e^{\pi \alpha_j} -M^2\alpha_j^2}{(e^{\pi \alpha_j}-1)^{2} \alpha_j^{3/2}} \Bigg{]}.
\end{eqnarray}
Thus the equation \eqref{appendix_Dirichlet_small_M_1} becomes
\begin{eqnarray}
 \label{appendix_Dirichlet_small_M_2}
 \begin{aligned}
  {\cal I}_j \approx \frac{1}{16 \alpha_j^3} \Bigg{[} & 8\pi - 4\pi \Big{[}2+M\alpha_j(2+M\alpha_j)\Big{]}e^{-M\alpha_j} + 
  8M^3\alpha_j^2K_1(M\alpha_j) + \\ & + 24M^2\alpha_j K_2(M\alpha_j) - 8\pi \text{coth}\Big{(}\frac{\pi \alpha_j}{2}\Big{)}
  + \pi^2 \alpha_j \Big{[} M^2\alpha_j^2 - 4 - \\ &- 2\pi \alpha_j^2 \text{coth}\Big{(}\frac{\pi \alpha_j}{2}\Big{)}
  \Big{]}\text{csch}^2\Big{(}\frac{\pi \alpha_j}{2}\Big{)} \Bigg{]}.
 \end{aligned}
\end{eqnarray}

Considering the parameter $\alpha_j = j\sigma$ large, we get
\begin{eqnarray}
 \label{appendix_Dirichlet_small_M_3}
 \begin{aligned}
  {\cal I}_j \approx \frac{M^3}{2j\sigma}K_1(jM\sigma) + \frac{3M^2}{2j^2\sigma^2}K_2(jM\sigma) - \frac{\pi[2+jM\sigma(2+
  jM\sigma)]e^{-jM\sigma}}{4j^3\sigma^3}.
 \end{aligned}
\end{eqnarray}
At this point, two distinct limits can be analyzed: $M \sigma > 1$ and $M\sigma < 1$.
\begin{itemize}
 \item For case $M\sigma > 1$: In this case  we can use the asymptotic expression for the modified Bessel of second kind function for large argument \eqref{large_Bessel}, obtaining 
 \begin{eqnarray}
 \label{appendix_Dirichlet_small_M_4}
 \begin{aligned}
  {\cal I}_j \approx \sqrt{\frac{\pi M^5}{8\sigma^3}} \frac{e^{-jM\sigma}}{j^{3/2}} + \sqrt{\frac{9\pi M^3}{8\sigma^5}} 
  \frac{e^{-j M\sigma}}{j^{5/2}} -\frac{\pi[2+jM\sigma(2+jM\sigma)]e^{-jM\sigma}}{4j^3\sigma^3}.
  \end{aligned}
 \end{eqnarray}
 Therefore, the dominant contribution in the sum of $j$ of the expression above is given by the term with $j=1$.

\end{itemize}
 
\begin{itemize}
 \item For case $M\sigma < 1$, the sum of \eqref{appendix_Dirichlet_small_M_3} is given by:
 \begin{eqnarray}
 \label{appendix_Dirichlet_small_M_5}
  {\cal S} &\approx &\sum_{n=1}^{\infty}{\cal I}_j \approx \frac{M^3}{2\sigma} \sum_{j=1}^{\infty} \frac{K_1(j M \sigma)}{j} + 
  \frac{3M^2}{2\sigma^2}\sum_{j=1}^{\infty} \frac{K_2(j M \sigma)}{j^2} \nonumber\\
   & -& \frac{\pi}{4\sigma^3} \sum_{j=1}^{\infty} \frac{[2+jM\sigma(2+
  jM\sigma)]e^{-jM\sigma}}{j^3}.
\end{eqnarray}
Unfortunately there is no closed expression for the summations above involving the Bessel function, $K_\nu(z)$. So, we need an approximated expression for them. We know that the Bessel function of second kind is a decreasing function. In fact for large argument this function presents a exponential decay. Because we are using $M\sigma<<1$, the argument of both Bessel functions become of the order of unity for very large value of $j$; however, this same number appears in the denominators, which indicates that the ration $\frac{K_n(jM\sigma)}{j^n}$ becomes very small. With this restriction, we can show that the sums above can be approximated by considering only the firsts terms of the expansions of the corresponding Bessel functions for small arguments. Accepting this fact we can write:
\begin{eqnarray}
 \label{appendix_Dirichlet_small_M_approx}
 \begin{aligned}
 \sum_{j=1}^{\infty} \frac{K_1(j M \sigma)}{j} \approx \frac{\pi^2}{6M \sigma}\ \ \ \text{and} \ \ \ \sum_{j=1}^{\infty} 
 \frac{K_2(j M \sigma)}{j^2} \approx \frac{4 \pi ^4-15 \pi ^2 M^2 \sigma ^2}{180 M^2 \sigma ^2}.
 \end{aligned}
\end{eqnarray}
So, we found
 \begin{eqnarray}
 \label{Aappendix_Dirichlet_small_M_6}
  {\cal S} &\approx& \frac{\pi^2 M^2}{12 \sigma^2} + \frac{12\pi^4M^2-45\pi^2M^4\sigma^2}{360M^2\sigma^4}+ \frac{1}{4\sigma^3}
  \Bigg{[} \pi M^2 \sigma^2 \text{Ln}\Big{(}1-e^{-M\sigma}\Big{)}\nonumber\\
   & -&2\pi M\sigma \text{Li}_{2}\Big{(}e^{-M\sigma}\Big{)} - 
  2\pi \text{Li}_{3}\Big{(}e^{-M\sigma}\Big{)} \Bigg{]}  \  ,
\end{eqnarray}
where $Li_n(z)$ represents the general polylogarithm function \cite{abramowitz:1966}.
\end{itemize}

In oder to justify our approximation, Eq. \eqref{appendix_Dirichlet_small_M_approx}, exhibit numerically, in figure $2$, the behavior of the summations of the Bessel functions of second kind, comparing with their corresponding approximated expressions given in \eqref{appendix_Dirichlet_small_M_approx}. We can see in both panels a very good agreement between the exact functions, $\sum_j\frac{K_1(jM\sigma)}{j}$ and $\sum_j\frac{K_2(jM\sigma)}{j^2}$, with their corresponding approximations.
\begin{figure}[tbph]
	\begin{center}
		\begin{tabular}{cc}
			\epsfig{figure=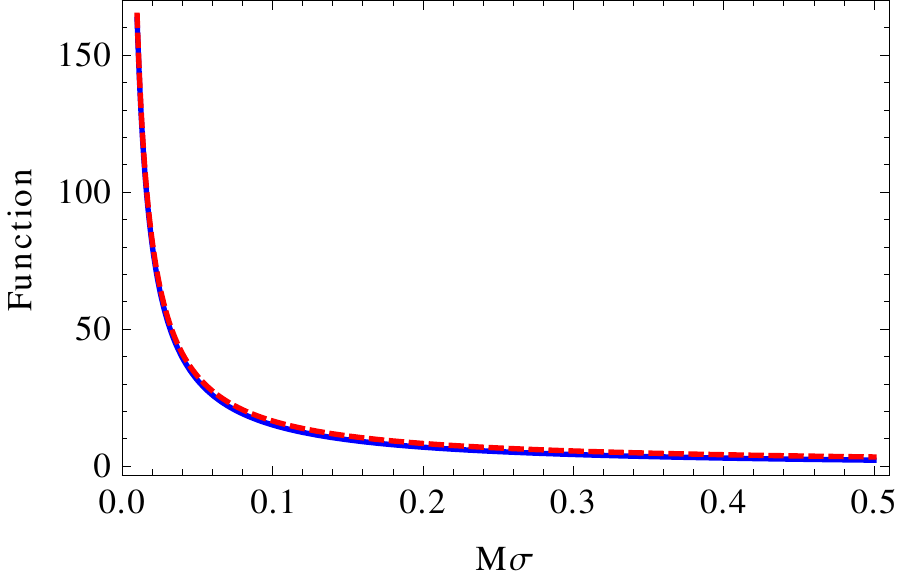,width=7.cm,height=6.cm} & \quad %
			\epsfig{figure=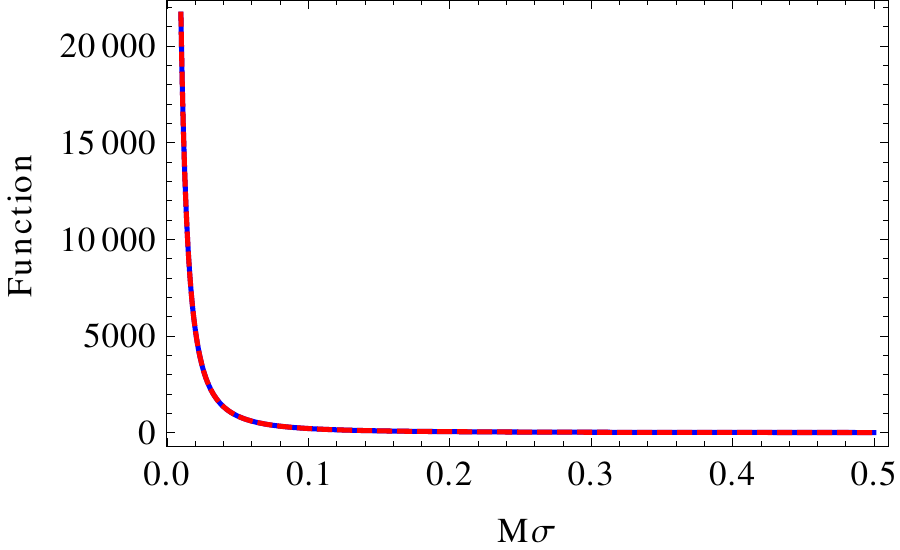,width=7.cm,height=6.cm}%
		\end{tabular}%
	\end{center}
	\caption{In the left panel we exhibit the behavior of the summation involving the modified Bessel function $K_1(jM\sigma)/j$ in blue line, and the behavior of the function $\pi^2/(6M\sigma)$ in red  dashed line as function of $M\sigma$. In the right panel the same for the Bessel function $K_2(jM\sigma)/j^2$ in blue line and $\frac{4 \pi ^4-15 \pi ^2 M^2 \sigma ^2}{180 M^2 \sigma ^2}$ in dashed line as function of $M\sigma$.}
\end{figure}

 Now, taking the limit $M\sigma<<1$, the expression \eqref{Aappendix_Dirichlet_small_M_6}, becomes
  \begin{eqnarray}
  \label{appendix_Dirichlet_small_M_t}
  \begin{aligned}
  {\cal S} \approx \frac{\pi  \left(15 M^2 \sigma ^3-5 \pi  M^2 \sigma ^2-60 \sigma \zeta (3)+4 \pi ^3\right)}{120 \sigma ^4} .
  \end{aligned}
 \end{eqnarray}

\section{Calculation of integral II}
\label{appB}
Now we consider the integral that appears in \eqref{Mista_energy_Casimir_x_2}:
\begin{eqnarray}
 \label{appendix_B_01}
 \begin{aligned}
  {\cal I}_j & = \int_M^{\infty} dz z^3 \int_1^{\infty} dy \frac{\sqrt{y^2-1}}{e^{2zy}+1} \cos \Big{(} \alpha_j z 
  \sqrt{y^2-1} \Big{)} \\
  & = \frac{\partial}{\partial \alpha_j} \int_M^{\infty} dz z^2 \int_1^{\infty} dy \frac{1}{e^{2zy}+1} \sin \Big{(}
  \alpha_j z \sqrt{y^2-1} \Big{)} .
 \end{aligned}
\end{eqnarray}
Using the relation below,
\begin{eqnarray}
 \label{appendix_B_02}
 \begin{aligned}
  \frac{1}{e^z+1} = - \sum_{n=1}^{\infty} (-1)^n e^{-nz},
 \end{aligned}
\end{eqnarray}
the expression \eqref{appendix_B_01} can be rewritten by
\begin{eqnarray}
 \label{appendix_B_03}
 \begin{aligned}
  {\cal I}_j & = - \frac{\partial}{\partial \alpha_j} \sum_{n=1}^{\infty} (-1)^n \int_M^{\infty} dz z^2 \int_1^{\infty} dy
  e^{-2nzy} \sin \Big{(} \alpha_j z \sqrt{y^2-1} \Big{)}  \\
  & = - \frac{\partial}{\partial \alpha_j} \Bigg{[} \alpha_j \sum_{n=1}^{\infty} \frac{(-1)^n}{\sqrt{\alpha_j^2+4n^2}} 
  \int_M^{\infty} dz z^2 K_1 \Big{(}z \sqrt{\alpha_j^2+4n^2}\Big{)} \Bigg{]}  \  ..
 \end{aligned}
\end{eqnarray}
where we have used the integral representation for the Bessel function, $K_\mu(z)$, given in \cite{prudnikov:1986}. Performing the change of variable $x = z \sqrt{\alpha_j^2+4n^2}$, we get
\begin{eqnarray}
 \label{appendix_B_04}
 \begin{aligned}
  {\cal I}_j = - \frac{\partial}{\partial \alpha_j} \Bigg{[} \alpha_j \sum_{n=1}^{\infty} \frac{(-1)^n}{(\alpha_j^2+4n^2)^2} 
  \int_{M\sqrt{\alpha_j^2+4n^2}}^{\infty} dx x^2 K_1(x) \Bigg{]} ,
 \end{aligned}
\end{eqnarray}
therefore, using \eqref{recorrence_relation} we get
\begin{eqnarray}
 \label{Integral_2_3}
 \begin{aligned}
  {\cal I}_j & = \frac{\partial}{\partial \alpha_j} \Bigg{[} \alpha_j \sum_{n=1}^{\infty} \frac{(-1)^n}{(\alpha_j^2+4n^2)^2} 
  \int_{M\sqrt{\alpha_j^2+4n^2}}^{\infty} \frac{\partial}{\partial x} \Big{[} x^2 K_2(x) \Big{]} dx \Bigg{]} \\
  & = - M^2 \frac{\partial}{\partial \alpha_j} \Bigg{[} \alpha_j \sum_{n=1}^{\infty} \frac{(-1)^n}{\alpha_j^2+4n^2} 
  K_2 \Big{(} M \sqrt{\alpha_j^2+4n^2} \Big{)} \Bigg{]}.
 \end{aligned}
\end{eqnarray}
Here we also will consider, as before, three different cases: $M \gg 1$, $M = 0$ and $M \ll 1$.

\begin{itemize}
 \item For case $M \gg 1$, the dominant contribution is $n=1$:
 \begin{eqnarray}
  \label{Integral_2_M_g}
  \begin{aligned}
   {\cal I}_j & \approx M^2 \frac{\partial}{\partial \alpha_j} \Bigg{[} \frac{\alpha_j}{\alpha_j^2+4} K_2 \Big{(} M 
   \sqrt{\alpha_j^2+4} \Big{)} \Bigg{]} .
  \end{aligned}
 \end{eqnarray}
 In this case, we can use the asymptotic expression for the modified Bessel function of second kind for large arguments \eqref{large_Bessel}. Thus we get
 \begin{eqnarray}
  \label{Integral_2_M_g_2}
  \begin{aligned}
   {\cal I}_j & \approx \sqrt{\frac{\pi M^3}{2}} \frac{\partial}{\partial \alpha_j} \Bigg{[} \frac{\alpha_j e^{-M\sqrt{\alpha_j+4}}}
   {(\alpha_j^2+4)^{5/4}} \Bigg{]} \\
   & \approx \sqrt{\frac{\pi M^3}{8}} \Bigg{[} \frac{8-\alpha_j^2\Big{(}3+2M\sqrt{\alpha_j^2+4}\Big{)}}{(\alpha_j^2+4)^{9/4}} 
   \Bigg{]} e^{-M\sqrt{\alpha_j^2+4}} .
  \end{aligned}
 \end{eqnarray}
 In limit that $\alpha_j \gg 1$, we get
 \begin{eqnarray}
  \label{Integral_2_M_g_3}
  \begin{aligned}
   {\cal I}_j & \approx - \sqrt{\frac{\pi M^5}{2}} \frac{e^{-M \alpha_j}}{\alpha_j^{3/2}} .
   \end{aligned}
 \end{eqnarray}
 \item For case $M = 0$:  In this case, we can use the asymptotic Bessel expression to small arguments \eqref{small_Besseal}. Thus, we find
 \begin{eqnarray}
  \label{Integral_2_M_0_1}
  \begin{aligned}
   {\cal I}_j = - \frac{1}{8} \frac{\partial}{\partial \alpha_j} \Bigg{[} \alpha_j \sum_{n=1}^{\infty} \frac{(-1)^n}
   {(n^2+\alpha_j^2/4)^2} \Bigg{]} = \frac{1}{4} \frac{\partial^2}{\partial \alpha_j^2} \sum_{n=1}^{\infty} \frac{(-1)^n}
   {n^2+\alpha_j^2/4} .
  \end{aligned}
 \end{eqnarray}
For this case,
\begin{eqnarray}
 \label{sum_2}
 \begin{aligned}
  \sum_{n=1}^{\infty} \frac{(-1)^n}{n^2+\alpha_j^2/4} = \frac{\pi \alpha_j \text{csch}\Big{(}\frac{\pi \alpha_j}{2}\Big{)} - 2}
  {\alpha_j^2} .
 \end{aligned}
\end{eqnarray}

Substituting \eqref{sum_2} into \eqref{Integral_2_M_0_1} and making the differentiation, we get
\begin{eqnarray}
 \label{Integral_2_M_0_2}
  {\cal I}_j &=& \frac{1}{16 \alpha_j^4} \Bigg{[} - 48 + \frac{1}{2}\pi \alpha_j \text{csch}^3\Big{(}\frac{\pi \alpha_j}{2}\Big{)}
   \Big{(}-8+3\pi^2\alpha_j^2+(8+\pi^2\alpha_j^2)\text{cosh}(\pi \alpha_j) \nonumber\\
  &+&4\pi \alpha_j \text{sinh}(\pi \alpha_j)\Big{)}\Bigg{]} \  . 
\end{eqnarray}
Considering the $\alpha_j \gg 1$, we get
\begin{eqnarray}
 \label{Integral_2_M_0_3}
 \begin{aligned}
  {\cal I}_j \approx - \frac{3}{\alpha_j^4}.
 \end{aligned}
\end{eqnarray}
 \item For case $M \ll 1$: For this case the expression \eqref{Integral_2_3} is not convenient to analyze this limit. So, we adopt here the same procedure as presented in Appendix \ref{appA}. Using the definition \eqref{function_f}, the expression  \eqref{Integral_2_3} can be written as, 
\begin{eqnarray}
 \label{appendix_Mixed_small_1}
 \begin{aligned}
  {\cal I}_j = - M^4 \frac{\partial}{\partial \alpha_j} \Bigg{[} \frac{\alpha_j}{2} \sum_{n=-\infty}^{\infty} (-1)^n
  f_2\Big{(}M\sqrt{\alpha_j^2+4n^2}\Big{)} - \frac{\alpha_j}{2} f_2\Big{(}M\alpha_j\Big{)} \Bigg{]} .
 \end{aligned}
\end{eqnarray}
Using the relation of sum \eqref{sum_relation}, with $\alpha=\pi$, the sum in the above equation reads,
\begin{eqnarray}
\label{sum_sigma}
 \begin{aligned}
 \Sigma_j & = \sum_{n=-\infty}^{\infty} (-1)^n f_2\Big{(}M\sqrt{\alpha_j^2+4n^2}\Big{)} \\
 & = \frac{\sqrt{2\pi}}{2M^4} \sum_{n=-\infty}^{\infty} \Big{[}\pi^2\Big{(}n+1/2\Big{)}^2+M^2\Big{]}^{3/2} f_{3/2}\Big{(}
 \alpha_j\sqrt{\pi^2\Big{(}n+1/2\Big{)}^2+M^2}\Big{)} .
 \end{aligned}
\end{eqnarray}
Substituting \eqref{sum_sigma} into \eqref{appendix_Mixed_small_1} we get
\begin{eqnarray}
 \label{appendix_Mixed_small_2}
 \begin{aligned}
  {\cal I}_j = - \frac{\partial}{\partial \alpha_j} & \Bigg{[} \frac{\sqrt{2\pi}}{4 \alpha_j^{1/2}} \sum_{n=-\infty}^{\infty}
  \Big{[}\pi^2\Big{(}n+1/2\Big{)}^2+M^2\Big{]}^{3/4} K_{3/2}\Big{(}\alpha_j\sqrt{\pi^2\Big{(}n+1/2\Big{)}^2+M^2}\Big{)} \\ 
  & - \frac{M^2}{2\alpha_j} K_2\Big{(}M\alpha_j\Big{)} \Bigg{]} .
 \end{aligned}
\end{eqnarray}

Because $M \ll 1$, we can use the approximate expression below
\begin{eqnarray}
 \begin{aligned}
  & \sum_{n=-\infty}^{\infty} [\pi^2(n+1/2)^2 + M^2]^{3/4} K_{3/2}(\alpha_j \sqrt{\pi^2(n+1/2)^2 + M^2}) \approx \\
  & \approx \sqrt{\frac{\pi}{2}} \frac{e^{\pi \alpha_j/2}}{(e^{\pi \alpha_j}-1)^2 \alpha_j^{3/2}} \Big{[} -2 + 2e^{\pi \alpha_j}
  +\pi \alpha_j + \pi \alpha_j e^{\pi \alpha_j} + M^2 \alpha_j^2 - M^2\alpha_j^2 e^{\pi \alpha_j} \Big{]} .
 \end{aligned}
\end{eqnarray}
Thus the equation \eqref{appendix_Mixed_small_2} becomes
\begin{eqnarray}
 \label{appendix_Dirichlet_small_M_3B}
 \begin{aligned}
  {\cal I}_j \approx \frac{1}{4 \alpha_j^3} & \Bigg{[} -2M^3\alpha_j^2K_1(M\alpha_j)-6M^2\alpha_jK_2(M\alpha_j)+
  \frac{\pi e^{3\pi\alpha_j/2}}{(e^{\pi \alpha_j}-1)^3} \Big{(} -8+3\pi^2\alpha_j^2 \\ 
  &+(8+\pi^2\alpha_j^2)\text{cosh}(\pi \alpha_j) +\pi \alpha_j(4-M^2\alpha_j^2)\text{sinh}(\pi \alpha_j) \Big{)}\Bigg{]} .
 \end{aligned}
\end{eqnarray}

Considering the parameter $\alpha_j = j\sigma$ large, we get
\begin{eqnarray}
 \label{appendix_Mixed_small_4}
 \begin{aligned}
  {\cal I}_j & \approx - \frac{M^3}{2\sigma}\frac{K_1(jM\sigma)}{j} - \frac{3M^2}{2\sigma^2}\frac{K_2(jM\sigma)}{j^2} .
 \end{aligned}
\end{eqnarray}
At this point, two distinct limits can be analyzed: $M \sigma > 1$ and $M\sigma < 1$.
\begin{itemize}
 \item For case $M\sigma > 1$: With this restriction we can use the asymptotic expression for the modified Bessel function, $K_\mu(z)$, for large argument \eqref{large_Bessel}. 
 Therefore, we get
 \begin{eqnarray}
 \label{appendix_Mixed_small_5}
 \begin{aligned}
  {\cal I}_j \approx - \sqrt{\frac{\pi M^5}{8 \sigma^3}} \frac{e^{-jM\sigma}}{j^{3/2}} - \sqrt{\frac{9\pi M^3}{8 \sigma^5}} 
  \frac{e^{-jM\sigma}}{j^{5/2}} .
  \end{aligned}
 \end{eqnarray}
 The dominant contribution in the sum over $j$ is given by the term with $j=1$.
 \item For case $M\sigma < 1$, the sum of \eqref{appendix_Mixed_small_4} is given by:
 \begin{eqnarray}
 \label{appendix_Dirichlet_small_7}
 \begin{aligned}
  {\cal S} \approx \sum_{j=1}^{\infty}{\cal I}_j \approx - \frac{M^3}{2\sigma} \sum_{j=1}^{\infty} \frac{K_1(jM\sigma)}{j} 
  - \frac{3M^2}{2\sigma^2} \sum_{j=1}^{\infty} \frac{K_2(jM\sigma)}{j^2} .
 \end{aligned}
\end{eqnarray}
Using the approximation \eqref{appendix_Dirichlet_small_M_approx}, we get
 \begin{eqnarray}
 \label{appendix_Dirichlet_small_M_6}
 \begin{aligned}
  {\cal S} \approx \frac{5 \pi ^2 M^2 \sigma ^2-4 \pi^4}{120 \sigma ^4} .
 \end{aligned}
\end{eqnarray}
\end{itemize}

\end{itemize}


\end{document}